\documentclass[12pt]{article}
\usepackage{geometry}
\usepackage{cite}
\usepackage{hyperref}
\usepackage{amsmath}
\usepackage{amssymb}
\usepackage{graphicx}
\usepackage[FIGTOPCAP]{subfigure}
\usepackage{wrapfig}
\usepackage{soul}

\usepackage{bbm}
\usepackage{caption}
\usepackage[T1]{fontenc}
\usepackage{capt-of}
\usepackage[font=small,labelfont=bf,tableposition=top]{caption}
\DeclareCaptionLabelFormat{andtable}{#1~#2  \&  \tablename~\thetable}

\usepackage{floatrow}
\usepackage[T1]{fontenc}
\usepackage[utf8]{inputenc}
\usepackage[font=small,labelfont=bf]{caption}
\newfloatcommand{capbtabbox}{table}[][\FBwidth]
\usepackage[table]{xcolor}
\usepackage{graphics}

\usepackage[amssymb]{SIunits}
\usepackage{stackengine}
\def\delequal{\mathrel{\stackon[1pt]{=}{$\scriptscriptstyle\Delta$}}}
\usepackage{array}
\usepackage{url}
\begin{document}

%------------------ Title/Authors/institutions/others ------------------------------
\vspace{-5mm}
%\title{Bayesian Image Quality Transfer with CNNs: Exploring Uncertainty in dMRI Super-Resolution}
\title{ Uncertainty Quantification in Deep Learning \\for Safer Neuroimage Enhancement}
%\title{Deep Uncertainty Quantification \\for Neuroimage Enhancement}

%\author{ Ryutaro~Tanno, Daniel~Worrall, Enrico~Kaden, Aurobrata~Ghosh,
%\\Felix Bragman, Zach Eaton-Rosen, Stefano B. Blumberg,
%\\ Francesco Grussu,  Stamatios~N.~Sotiropoulos, Antonio~Criminisi,
%\\  Daniel~C.~Alexander }

\author{
	Ryutaro Tanno$^{1}$, Daniel E. Worrall$^{2}$, Enrico Kaden$^{1}$,\\
Aurobrata Ghosh$^{1}$, Francesco Grussu$^{1,3}$, Alberto Bizzi$^{4}$,  \\
Stamatios N. Sotiropoulos$^{5,6}$, Antonio Criminisi$^{7}$, Daniel C. Alexander$^{1}$
}

\date{%
	\small
	$^1$Centre for Medical Image Computing and Dept. Computer Science, UCL, \\
	Gower Street, London, UK, WC1E 6BT. \\
	$^2$Machine Learning Lab, University of Amsterdam, The Netherlands.\\
	$^3$Institute of Neurology, Faculty of Brain Sciences, UCL, UK. \\
	$^4$Neuroradiology Unit, Foundation IRCCS Carlo Besta Neurological Institute, Milan, Italy. \\
	$^5$Sir Peter Mansfield Imaging Centre, School of Medicine and \\
	NIHR Biomedical Research Centre, University of Nottingham, UK.\\
	$^6$Wellcome Centre for Integrative Neuroimaging, University of Oxford, UK.\\
	$^7$Microsoft Research Cambridge, UK. \\
	\vspace{2mm}
	Correspondence: \href{mailto:r.tanno@cs.ucl.ac.uk}{\url{r.tanno@cs.ucl.ac.uk}}
}

\maketitle 
\markboth{Journal of \LaTeX\ Class Files,~Vol.~14, No.~8, August~2015}%
{Shell \MakeLowercase{\textit{\emph{et al.}}}: Bare Demo of IEEEtran.cls for IEEE Journals}



\maketitle

%------------------ Abstract ------------------------------------
\begin{abstract}
	Deep learning (DL) has shown great potential in medical image enhancement problems, such as super-resolution or image synthesis. However, to date little consideration has been given to uncertainty quantification over the output image. Here we introduce methods to characterise different components of uncertainty in such problems and demonstrate the ideas using diffusion MRI super-resolution.  Specifically, we propose to account for \textit{intrinsic uncertainty} through a heteroscedastic noise model and for \textit{parameter uncertainty} through approximate Bayesian inference, and integrate the two to quantify \textit{predictive uncertainty} over the output image. Moreover, we introduce a method to propagate the predictive uncertainty on a multi-channelled image to derived scalar parameters, and separately quantify the effects of intrinsic and parameter uncertainty therein. The methods are evaluated for super-resolution of two different signal representations of diffusion MR images---Diffusion Tensor images and Mean Apparent Propagator MRI---and their derived quantities such as mean diffusivity and fractional anisotropy, on multiple datasets of both healthy and pathological human brains. Results highlight three key potential benefits of uncertainty modelling for improving the safety of DL-based image enhancement systems. Firstly, incorporating uncertainty modelling improves the predictive performance even when test data departs from training data. Secondly, the predictive uncertainty highly correlates with reconstruction errors, and is therefore capable of detecting predictive ``failures''. Results on both healthy subjects and patients with brain glioma or multiple sclerosis demonstrate that such an uncertainty measure enables subject-specific and voxel-wise risk assessment of the super-resolved images that can be accounted for in subsequent analysis. Thirdly, we show that the method for decomposing predictive uncertainty into its independent sources provides high-level ``explanations'' for the model performance by separately quantifying how much uncertainty arises from the inherent difficulty of the task or the limited training examples. The introduced concepts of uncertainty modelling extend naturally to many other imaging modalities and data enhancement applications.

\end{abstract}

%------------------------------------ Introduction  ------------------------------
\section{Introduction}
%A potential structure of intro below: 
%\textcolor{red}{Make sure that context and contributions are separated. You need a concrete and well-referenced context development. Also, make sure our high-level contributions are evident.}

% 2018-01-10
%In the last few years, deep learning techniques have permeated the field of medical image processing \cite{shen2017deep,litjens2017survey}. Modern research in automation of radiological tasks through deep learning have shown a great promise in an array of applications such as segmentation \cite{kamnitsas2017efficient}, detection \cite{roth2014new}, disease grading and classification \cite{araujo2017classification}. Another notable category of problems is data enhancement, which aims to increase both the quality and quantity of medical images for research and clinics, and come in different forms such as super-resolution \cite{oktay2016multi}, image synthesis \cite{nie2016estimating,kang2017deep}, denoising \cite{benou2017ensemble}, data harminization \cite{karayumak2018harmonizing}, reconstruction \cite{jin2017deep,schlemper2018deep,zhu2018image,yang2018dagan}, registration \cite{sokooti2017nonrigid} and quality control \cite{wu2017fuiqa,esses2018automated}. These advances have the potential to not only increase the quality and efficiency of radiological care, but also facilitate scientific discoveries in medical research. 

In the last few years, deep learning techniques have permeated the field of medical image processing \cite{shen2017deep,litjens2017survey}. Beyond the automation of existing radiological tasks--- e.g. segmentation \cite{kamnitsas2017efficient}, detection \cite{roth2014new}, disease grading and classification \cite{araujo2017classification}---deep learning has been applied to a diverse set of ``data enhancement'' problems. Data enhancement aims to improve the quality, the information content, or the quantity of medical images available for research and clinics by transforming images from one domain to another \cite{isola2017image}. Previous research has shown the efficacy of data enhancement in different forms such as super-resolution \cite{oktay2016multi,chen2018efficient,ravi2019adversarial}, image synthesis \cite{nie2016estimating,kang2017deep}, denoising \cite{benou2017ensemble,chen2017low}, data harmonisation \cite{karayumak2018harmonizing,tax2019cross} across scanners and protocols, reconstruction \cite{sun2016deep,jin2017deep,hammernik2018learning,schlemper2018deep,zhu2018image,yang2018dagan,yoon2019efficient}, registration \cite{sokooti2017nonrigid,balakrishnan2018unsupervised} and quality control \cite{wu2017fuiqa,esses2018automated}.  These advances have the potential not only to enhance the quality and efficiency of radiological care, but also facilitate scientific discoveries in medical research through increased volume and content of usable data. 

However,  most efforts in the development of data enhancement techniques have focused on improving the accuracy of deep learning algorithms, with little consideration of risk management. Blindly trusting the output of a given machine learning tool risks undetected failures e.g. spurious features and removal of structures \cite{cohen2018distribution}. In medical applications, images inform scientific conclusions in research, and diagnostic, prognostic and interventional decisions in clinics. Therefore, translation of current proofs of principle to such safety-critical applications demands mechanisms for quantifying the risks of failures i.e. quantification of uncertainty/confidence and explanation of its source \cite{begoli2019need}. 

\begin{figure}[t]
	\includegraphics[width=0.95\linewidth]{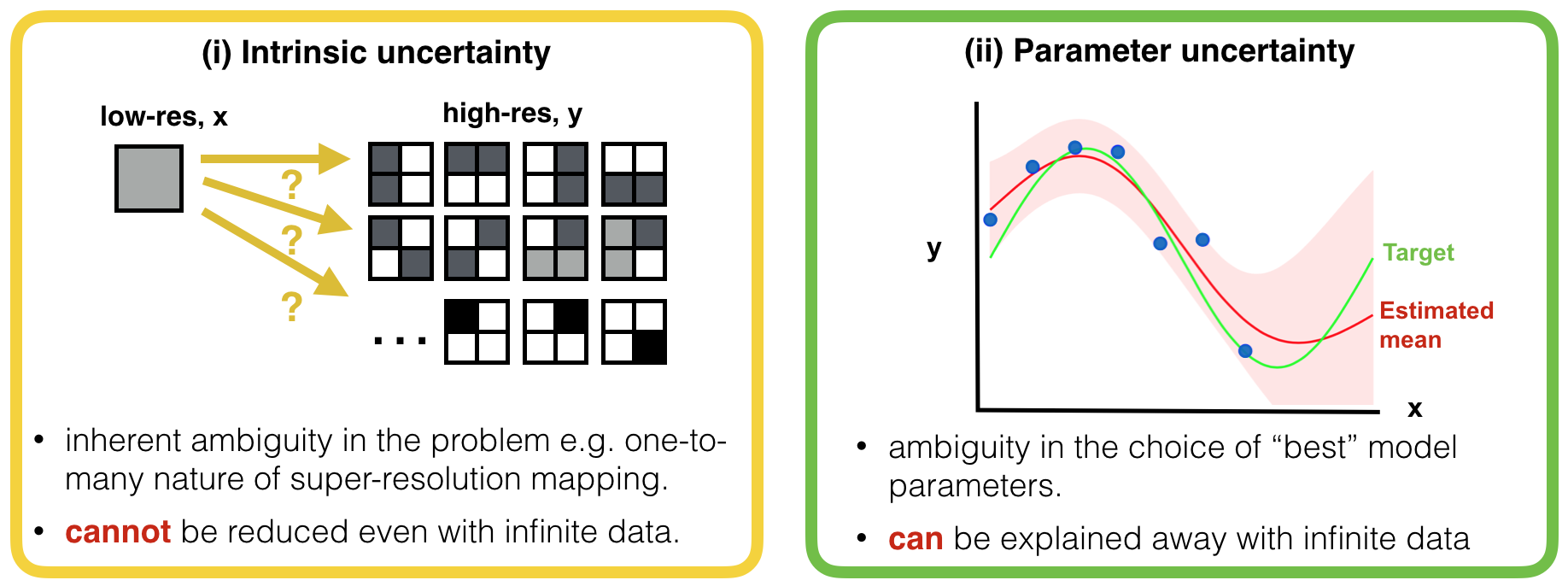}
	\centering	
%	\vspace{-2mm}
	\caption{Illustration of two different types of uncertainty \cite{hora1996aleatory}. Intrinsic uncertainty \cite{wang1996intrinsic} quantifies the degree of inherent ambiguity in the underlying problem. For example, in the case of super-resolution, there exist many possible high-resolution images $\textbf{y}$ that would get mapped onto the same low-resolution input $\textbf{x}$. Intrinsic uncertainty is irreducible with training data. On the other hand, the parameter uncertainty \cite{draper1995assessment} (a subtype of model uncertainty) arises from the finite training set. There exist more than one model that can explain the given training data equally well, and the parameter uncertainty quantifies the ambiguity in selecting the model parameters that best captures the target data-generating process. As illustrated in the figure on the right, parameter uncertainty decreases with more data; the green line shows the target function, the red line is the estimated mean, and the shaded region signifies the associated parameter uncertainty (standard deviation), which is higher in regions where we have fewer observations.} 
	\label{fig:uncertainty_types}
	%\vspace{-10pt}
\end{figure}

Predictive failures of deep learning systems, by and large, occur due to two reasons: i) the task itself is inherently ambiguous or ii) the learned model is not adequate to describe the data  \cite{hora1996aleatory,der2009aleatory,tanno2017bayesian,kendall2017uncertainties}, as illustrated in Fig.~\ref{fig:uncertainty_types}. The former stems from \emph{intrinsic uncertainty} \cite{wang1996intrinsic}, which describes ambiguity in the underlying data generating process  (e.g. presence of stochasticity such as measurement noise and intrinsic ill-posed nature of the problem), and cannot be alleviated by increasing available training data or model complexity\footnote{Intrinsic uncertainty is also known as \textit{aleaotoric} or statistical uncertainty.}. The latter is characterised by \textit{model uncertainty}\cite{draper1995assessment}, which describes ambiguity in model specification\footnote{Model uncertainty is a subclass of \textit{epistemic uncertainty} \cite{hora1996aleatory} which encompasses types of uncertainties that arise from lack of knowledge. }. Model uncertainty arises from a) \textit{parameter uncertainty}: ambiguity in fitting the model to the target mapping due to limited training data, or b) \textit{model bias}: errors due to insufficient flexibility of the model class (e.g. fitting a linear model to a sinusoidal process). These types of uncertainty can be reduced by collecting more data or specifying a different class of models. With the expressivity of deep neural networks, which are known to be universal approximators \cite{cybenko1989approximation} if sufficiently large, one might reasonably assume that the model bias is small enough to be discounted. Under this assumption, intrinsic and parameter uncertainty (Fig.~\ref{fig:uncertainty_types}) fully characterise the predictive failures of deep learning models. Therefore, accurate estimation of these uncertainties are needed and would potentially allow practioners to understand better the limits of the models, flag doubtful predictions, and highlight test cases that are not well represented in the training data.

%Fig.~\ref{fig:uncertainty_types} illustrates the differences between intrinsic and parameter uncertainty, the two key types of uncertainty that need to be considered to characterise the limits of deep learning models. 

% (Ryu): I think we should mention the previous work... 
%Kiureghian \emph{et al.}  \cite{der2009aleatory} first proposed such categorisation of uncertainty in statistical modelling, and adovocated its importance in the assessment of structural safety e.g. seismic collapse risk of buildings.  More recently, Tanno \emph{et al.}  \cite{tanno2017bayesian} extended this argment to deep learning systems and demonstarated the importance of modelling both intrinsic and parameter uncertainty to build more robust predictive models for medical imaging. Kendall \emph{et al.} \cite{kendall2017uncertainties} concurrently investigated the same problem in computer vision, suggesting its utility for safety-critical applications such as self-driving cars. 

%\textcolor{red}{What is unique about medical applications that puts particular demands on uncertainty quantification - is it different in any way to the challenge in computer vision?}

In this work, we introduce methods for capturing components of uncertainty in medical image enhancement systems based on deep learning. We propose to model intrinsic uncertainty through a input-dependent (heteroscedastic) noise model \cite{nix1994estimating} and parameter uncertainty through variational dropout \cite{kingma2015variational}. We then combine and propagate these two ``source'' uncertainties into a spatial map of \textit{predictive uncertainty} over the output image, which can be used to assess the output reliability on subject-specific and voxel-wise basis. Lastly, we propose a method to propagate the predictive uncertainty to arbitrary derived quantities of the output images, such as scalar indices that are commonly used for subsequent analysis, and decompose it into distinct components which separately quantify the contributions of intrinsic and parameter uncertainty. This paper demonstrates the benefits of these ideas to enhancing system safety within the context of Image Quality Transfer (IQT) \cite{alexander2014image,tanno2016bayesian,alexander2017image,blumberg2018deeper}, a data-enhancement framework for propagating information from rare or expensive high quality images to lower quality but more readily available images. We focus on the application of IQT to \textit{super-resolution} of diffusion magnetic resonance imaging (dMRI) scans, and evaluate the utility of uncertainty quantification in terms of three aspects; i) performance on unseen datasets; ii) safety assessment of system output; iii) explainability of failures. For two different types of diffusion signal representations, we evaluate the effects of uncertainty modeling on generalisation by measuring the predictive accuracy on unseen test subjects in the Human Connectome Project (HCP) dataset \cite{sotiropoulos2013advances} and the Lifespan dataset \cite{harms2018extending}. We additionally test the value of improved predictive performance in a downstream tractography application. We then test the capability of the predictive uncertainty map to indicate predictive errors and thus to detect potential failures on images of both healthy subjects and those in which pathologies unseen in the training data arise, specifically from glioma and multiple-sclerosis (MS) patients. Lastly, we perform the decomposition of predictive uncertainty on HCP subjects with benign abnormalities, and assess its potential value in gaining high-level interpretations of predictive performance.

\section{Related Works}
This section provides a review of related works under several different themes. We first review the development of learning-based image enhancement methods in medical imaging applications. We then discuss the recent advances made to model and quantify uncertainty in such image enhancement problems. Lastly, we describe the existing strands of research in uncertainty modelling for other medical imaging problems and fields of applications. 

Various forms of image enhancement can be cast as image transformation problems where the input image from one domain is mapped to an output image from another domain. Numerous recent methods have proposed to perform image transformation tasks as supervised regression of low quality against high quality image content. Alexander \emph{et al.} \cite{alexander2014image} proposed Image Quality Transfer (IQT), a general framework for supervised quality enhancement of medical images. They demonstrated the efficacy of their method through a random forest (RF) implementation of super-resolution (SR) of brain diffusion tensor images and estimation of advanced microstructure parameter maps from sparse measurements. More recently, deep learning, typically in the form of convolutional neural networks (CNNs), has shown additional promise in this kind of task. For example, Oktay \emph{et al.} \cite{oktay2016multi} proposed a CNN model to upsample a stack of 2D MRI cardiac volumes in the through-plane direction, where the SR mapping is learnt from 3D cardiac volumes of nearly isotropic voxels. This work was later extended by \cite{oktay2018anatomically} with the addition of global anatomical prior based on auto-encoder. Zhao \emph{et al.} \cite{zhao2018deep} proposed a solution to the same SR problem for brains that utilises the high frequency information in in-plane slices to super-resolve in the through-plane direction without requiring external training data. In addition, a range of different architectures of CNNs have been considered for SR of other modalities and anatomical structures such as structural MRI \cite{chen2018efficient} of brains, retinal fundus images \cite{mahapatra2017image} and computer tomography (CT) scans of chest \cite{yu2017computed}. Another problem of growing interest is image synthesis, which aims to synthesise an image of a different modality given the input image. Nie \emph{et al.} \cite{nie2018medical} employed a conditional generative adversarial network to synthesise CT from MRI with fine texture details whilst Wolterink \emph{et al.} \cite{wolterink2017deep} extended this idea using a CycleGAN \cite{zhu2017unpaired} to leverage the abundance of unpaired training sets of CT and MR scans. In \cite{bahrami2016convolutional}, a variant of CNN  was applied to predict 7T images from 3T MRI, where both contrast and resolution are enhanced. Another notable application is the harmonisation of diffusion MRIs \cite{karayumak2018harmonizing,tax2019cross,blumberg2018deeper,blumberg2019msp} where images acquired at different scanners or magnetic field strengths are mapped to the common reference image space to allow for joint analysis.  

Despite this advancement, all of these methods commit to a single prediction and lack a mechanism to communicate uncertainty in the output image.  In medical applications where images can ultimately inform life-and-death decisions, quantifying reliability of output is crucial. Tanno \emph{et al.} \cite{tanno2016bayesian} aimed to address this problem for supervised image enhancement for the first time by proposing a Bayesian variant of random forests to quantify uncertainty over predicted high-resolution MRI. They showed that the uncertainty measure correlates well with the accuracy and can highlight abnormality not represented in the training data. In our preliminary work \cite{tanno2017bayesian}, we made an initial attempt
to extend this approach with probabilistic deep-learning formulation, and showed that modelling different components of uncertainty---intrinsic and parameter uncertainty---allows one to build a more generalisable model and quantify predictive confidence. Kendall \emph{et al.} \cite{kendall2017uncertainties} concurrently investigated the same problem in computer vision, suggesting its utility for safety-critical applications such as self-driving cars. More recently, Hu \emph{et al.}\cite{hu2019uncertainty} extended these works in the context of medical image segmentation and proposed a mechanism to learn the intrinsic uncertainty in a supervised manner, when multiple labels are available. Dalca \emph{et al.} \cite{dalca2018unsupervised} proposed a CNN-based probabilistic model for diffeomorphic image registration with a learning algorithm based on variational inference, and demonstrated the state-of-the-art registration accuracy on established benchmarks while providing estimates of registration uncertainty. An alternative approach is ensembling where the variance of the predictions of multiple networks is used to quantify the predictive uncertainty \cite{lakshminarayanan2017simple}. Schlemper \emph{et al.} \cite{schlemper2018stochastic} proposed a novel combination of the cascaded CNN architecture and compressive sensing, equipped with a variant of ensemble techniques, which enabled robust reconstruction of highly undersampled cardiovascular diffusion MR images, and quantification of reconstruction uncertainty. Bragman \emph{et al.} \cite{bragman2018uncertainty} studied the value of uncertainty modelling for multi-task learning in the context of MR-only radiotherapy treatment planning where the synthetic CT image and the segmentation of organs at risk are simultaneously predicted from the input MRI image.  

We should also note that, although not the focus of this work, research on uncertainty modelling in deep learning techniques extend to other medical image processing tasks beyond data enhancement, such as segmentation, detection and classification. For example, Nair \emph{et al.}, \cite{nair2018exploring} demonstrated for lesion segmentation of multiple sclerosis that the voxel-wise uncertainty metrics can be used for quality control; by filtering out predictions with high uncertainty, the model could achieve higher lesion detection accuracy. A concurrent work by Eaton-Rosen \emph{et al.} \cite{eaton2018towards} showed for the task of brain tumour segmentation that the Monte Carlo (MC) sample variance from dropout \cite{gal2015dropout} can be calibrated to provide meaningful error bars over estimates of tumour volumes. Similarly,  \cite{roy2019bayesian} introduced ways to turn voxel-wise uncertainty score into structure-wise uncertainty metrics for brain parcellation task, and showed their values in performing more reliable group analysis. The uncertainty metric based on MC dropout has also shown promise in disease grading of retinal fundal images \cite{worrall2016automated,leibig2017leveraging}, and more recently an extension based on test-time augmentation was introduced by \cite{ayhan2018test}. An alternative approach is to train a model to predict uncertainty score directly; \cite{Raghu2018DirectUP} showed that this approach is more effective when opinions from multiple experts are available for each image. Koh \emph{et al.} \cite{kohl2018probabilistic} and Baumgartner \emph{et al.} \cite{PHiSeg2019Baumgartner} proposed methods to generate a set of diverse and plausible segmentation proposals on a given image, capturing more realistically the high inter-reader annotation variability, which is commonly observed in medical image segmentation tasks. Lastly, \cite{raykar2010learning,tanno2019learning} demonstrated for the classification of mammograms and cardiac ultra-sound images, respectively that modelling uncertainty and biases of individual annotators enables robust learning from noisy labels in the presence of large disagreement. 

However, within the context of medical image enhancement, these lines of research performed only limited validation of the quality and utility of uncertainty modelling. In this work, we formalise and extend the preliminary ideas in Tanno \emph{et al.} \cite{tanno2017bayesian} and provide a comprehensive set of experiments to evaluate the proposed uncertainty modelling techniques in a diverse set of datasets, which vary in demographics, scanner types, acquisition protocols or pathology. Moreover, with the exception of \cite{tanno2017bayesian}, none of the previous methods model different components of uncertainty, namely intrinsic and parameter uncertainty. Our method accounts for both, and provides conclusive evidence that this improves performance thanks to different regularisation effects. In addition, we propose a method to decompose predictive uncertainty over an arbitrary function of the output image (e.g. morphological measurements) into its sources, in order to provide a high-level explanation of model performance on the given input.

\section{Methods}
This section describes the methods for modelling different components of uncertainty that arise in data enhancement. Firstly, we provide an overview of Image Quality Transfer (IQT) which formulates data enhancement as a supervised learning problem. Secondly, using the IQT framework, we introduce methods to model \textit{intrinsic} and \textit{parameter uncertainty}, separately, focusing on the application of super-resolution. We then combine the two approaches and estimate the overall uncertainty over prediction (\textit{predictive uncertainty}) by approximating the variance of the predictive distribution (eq.~\eqref{eq:full_distribution}). Lastly, we propose a method for decomposing predictive uncertainty into its sources---intrinsic and parameter uncertainty---in an attempt to provide quantifiable explanations for the confidence on model output (eq.~(\ref{eq:variance_decomposition})).

\subsection{Background: Image Quality Transfer}
\begin{figure}[t]
	\includegraphics[width=0.8\linewidth]{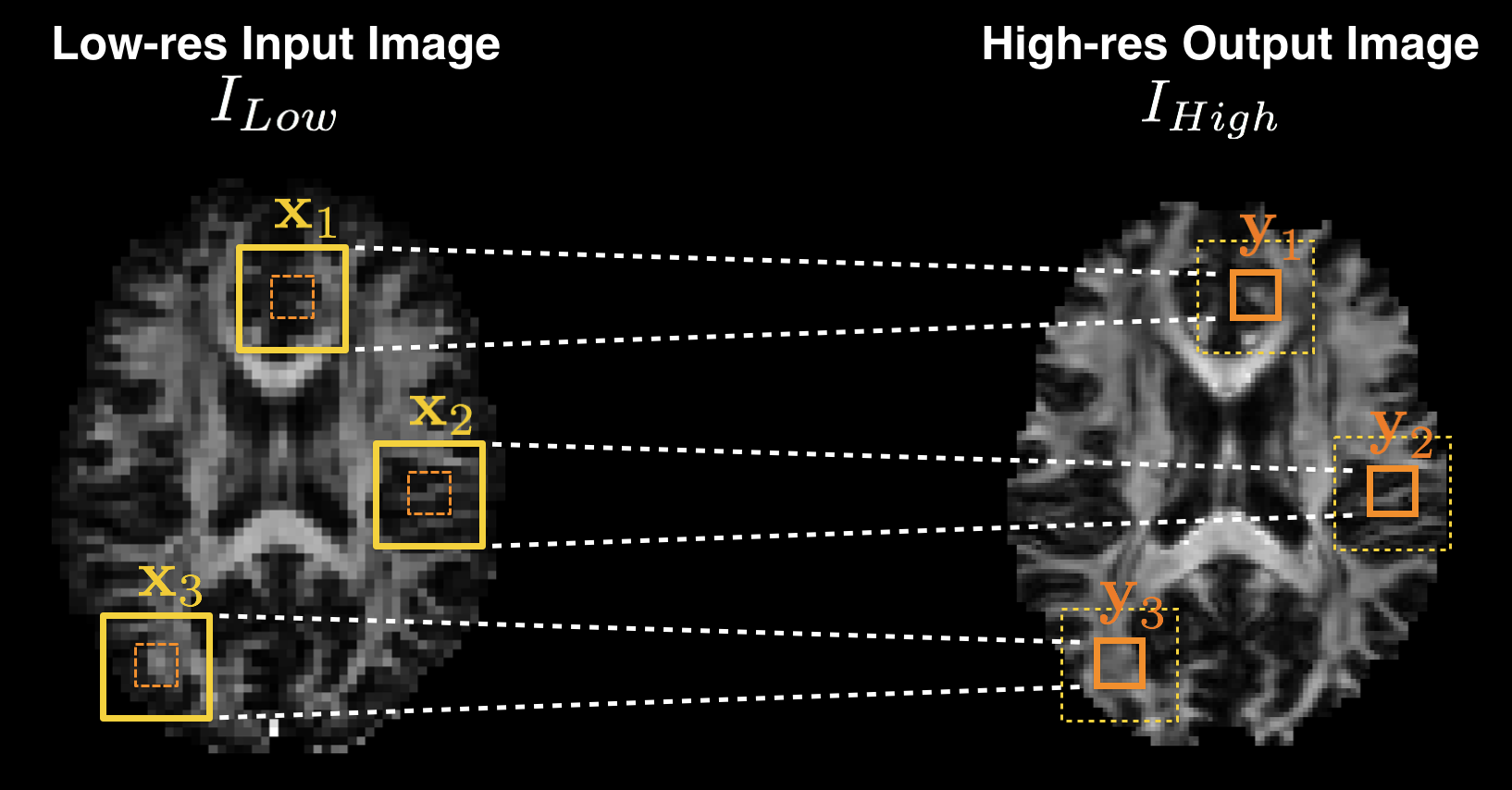}
	\centering	

	\caption{Illustration of the patch-wise regression in super-resolution application. The conditional distribution over the high quality image $p(I_{High}|I_{Low})$ is assumed to factorise over local neighbourhoods $\{(\mathbf{x}_i, \mathbf{y}_i)\}_{i}$. In this case, for each input subvolume $\mathbf{x}_{i}$ (in yellow), the high resolution version of the smaller centrally located neighbourhood, $\mathbf{y}_{i}$ (in orange) is regressed. } 
	\label{fig:iqt_illustration}
	
	%\vspace{-10pt}
\end{figure}

Alexander \emph{et al.} \cite{alexander2014image} proposed Image Quality Transfer (IQT), the first supervised learning based framework for data enhancement of medical images, and here we survey its general formulation which forms the testing ground of this work. IQT performs data enhancement via regression of low quality against high quality image content. In order to overcome the memory demands of processing 3-dimensional medical images, along with other subsequent work such as \cite{yang2016fast,oktay2016multi,bahrami2016convolutional,oktay2018anatomically}, IQT assumes factorisability over local neighbourhoods (also called patches) and models the conditional distribution of high-quality image $I_{High}$ given the corresponding low-quality input $I_{Low}$ as: 
\begin{equation}
p(I_{High}|I_{Low}) = \prod_{i\in \mathcal{S}} p(\mathbf{y}_{i}|\mathbf{x}_{i})
\end{equation}
where $\{\textbf{y}_{i}\}_{i \in \mathcal{S}}$ is a set of disjoint high-quality subvolumes with $\mathcal{S}$ denoting the set of their indices, which together constitute the whole image $I_{High}$, while $\{\textbf{x}_{i}\}_{i \in \mathcal{S}}$ is a set of potentially overlapping low-quality subvolumes, each of which contains and is spatially larger than the corresponding $\textbf{y}_{i}$, as illustrated in Fig.~\ref{fig:iqt_illustration}. Here we assume that each local neighbourhood is a cubic sub-volume. The locality assumption reduces the problem of learning $p(I_{High}|I_{Low})$ to the much less memory intensive problem of learning $p(\mathbf{y}|\mathbf{x})$. In other words, IQT formulates the data enhancement task as a patch-wise regression where an input low-quality image $I_{Low}$ is split into smaller overlapping sub-volumes  $\{\textbf{x}_{i}\}_{i \in \mathcal{S}}$ and the corresponding non-overlapping high-quality sub-volumes $\{\textbf{y}_{i}\}_{i \in \mathcal{S}}$ are independently predicted according to the patch regressor $p(\mathbf{y}|\mathbf{x})$. The final prediction for the 3D high-quality volume $I_{high}$ is constructed by tesellating the output patches $\{\textbf{y}_{i}\}_{i \in \mathcal{S}}$. 

The original implementation of IQT \cite{alexander2014image,alexander2017image,tanno2016bayesian} employed a variant of random forests (RFs) to model $p(\mathbf{y}|\mathbf{x})$ while more recent \cite{yang2016fast,oktay2016multi,bahrami2016convolutional,oktay2018anatomically} approaches use variants of convolutional neural networks (CNNs). Either way, the machine learning algorithm is trained on pairs of high-quality and low-quality patches $\mathcal{D}=\{(\mathbf{x}_i,\mathbf{y}_i)\}_{i=1}^N$ extracted from a set of image volumes, and is used to perform the data-enhancement task of interest.  Typically, such patch pairs $\mathcal{D}$ are synthesised by down-sampling a collection of high quality images to approximate their counterparts in a particular low-quality scenario \cite{alexander2014image,oktay2016multi}. In this work, we focus on the task of super-resolution (SR) where the spatial resolution of $I_{high}$ is higher than the input image $I_{low}$. 

\subsection{Baseline Super-Resolution Model: 3D-ESPCN} \label{sec:baseline}
As the baseline architecture for modelling $p(\mathbf{y}|\mathbf{x})$, we adapt efficient subpixel-shifted convolutional network (ESPCN) \cite{shi2016real} to 3D data. ESPCN is a recently proposed method with the capacity to perform real-time per-frame SR of videos while retaining high accuracy on 2D natural images.  We have chosen to base on this architecture for its simplicity and computational performance. Most CNN-based SR techniques first up-sample a low-resolution input image (e.g. through bilinear interpolation\cite{dong2016image}, deconvolution\cite{oktay2016multi,mcdonagh2017context}, fractional-strided convolution\cite{johnson2016perceptual}, etc) and then refine the high-resolution estimate through a series of convolutions. These methods suffer from the fact that $(1)$ the up-sampling can be a lossy process and $(2)$ refinement in the high-resolution space has a higher computational cost than in the low-resolution space. By contrast, ESPCN performs convolutions in the low-resolution-space, upsampling afterwards. The reduced resolution of feature maps dramatically decreases the computational and memory costs, which is more pronounced in processing 3D data.

\begin{figure}[t]
	\includegraphics[width=\linewidth]{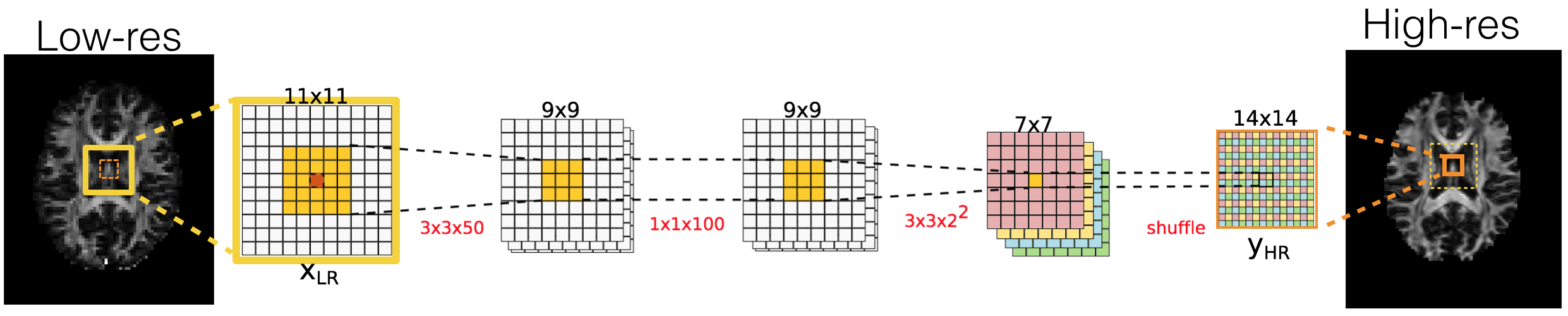}
	\centering	
	\caption{2D illustration of an example baseline network (ESPCN \cite{shi2016real}) with upsampling rate, $r=2$. The receptive field of the central $2^2$ pixels in the output patch is $5^2$ pixels in the input patch and
		 is shown in yellow. The shuffling operation at the end periodically rearranges the final feature maps from the low-resolution space into the high-resolution space.} 
	\label{fig:ESPCN}
	
	%\vspace{-10pt}
\end{figure}

More specifically the ESPCN is a fully convolutional network, with a special \emph{shuffling operation} on the output, which identifies individual feature channel dimensions with spatial locations in the high-resolution output. Fig.~\ref{fig:ESPCN} shows a 2D illustration of an example ESPCN when the fully convolutional part of the network consists of 3 convolutional layers, each followed by a ReLU, and the final layer has $cr^2$ feature maps where $r$ is the upsampling rate and $c$ is the number of channels in the output image (e.g. $6$ in the case of DT images). The shuffling operation takes the feature maps of shape $h\times w\times cr^2$ and remaps pixels from different channels into different spatial locations in the high-resolution output, producing a $rh\times rw\times c$ image, where $h$ and $w$ denote height and width of the pre-shuffling feature maps. This shuffling operation in 3D is given by $\mathcal{S}(F)_{i,j,k,c} =  F_{[i/r],[j/r],[k/r],(r^3-1)c + \text{mod}(i,r) + r\cdot \text{mod}(j,r) + r^3\cdot \text{mod}(k,r)}$ where $F$ is the pre-shuffled feature maps. The combined effects of the last convolution and shuffling is effectively a learned interpolation, and an efficient implementation of deconvolution layer \cite{zeiler2011adaptive} where the kernel size is divisible by the size of the stride  \cite{shi2016real}. Therefore, it is less susceptible to checker-board like artifacts commonly observed with deconvolution operations \cite{odena2016deconvolution}.

At test time, the prediction of higher resolution volume is performed through \textit{shift-and-stitch} operation. The network takes each subvolume $\mathbf{x}$ in a low-resolution image, and predicts the corresponding high-resolution sub-volume $\mathbf{y}$. By tessellating the predictions from appropriately shifted inputs $\mathbf{x}$, the whole high-resolution volume is reconstructed. With convolutions being local operations, each output voxel is only inferred from a local region in the input volume, and the spatial extent of this local connectivity is referred to as the \textit{receptive field}. For a given input subvolume, the network increases the resolution of the central voxel of each receptive field e.g. the central $2^3$ output voxels are estimated from the corresponding $5^3$ receptive field in the input volume, as coloured yellow in Fig.~\ref{fig:ESPCN}.

Given training pairs of high-resolution and low-resolution patches $\mathcal{D}=\{(\mathbf{x}_i,\mathbf{y}_i)\}_{i=1}^N$, we optimise the network parameters by minimising the sum of per-pixel mean-squared-error (MSE) between the ground truth $\mathbf{y}$ and the predicted high-resolution patch $\mu_{\theta}(\mathbf{x})$ over the training set. Here $\theta$ denotes all network parameters. This is equivalent to minimising the negative log likelihood (NLL) under the Gaussian noise model $p(\mathbf{y}|\mathbf{x},\mathbf{\theta}) = \mathcal{N}(\mathbf{y}; \mu_{\theta}(\mathbf{x}), \sigma^2I)$ with fixed isotropic variance $\sigma^2$. %Using the 3D-ESPCN as the basis, in the following sections we describe methods for modelling two different types of uncertainty, namely \textit{intrinsic} and \textit{parameter uncertainty} (Fig.~\ref{fig:uncertainty_types}). 
% Here, high-resolution patches are modelled as a deterministic function of low-resolution patches corrupted by fixed isotropic noise with variance $\sigma^2$. 

\subsection{Intrinsic Uncertainty and Heteroscedastic Noise Model \label{sec:hetero}} 
\begin{figure}
	%\vspace{-20pt}
	\includegraphics[width=\linewidth]{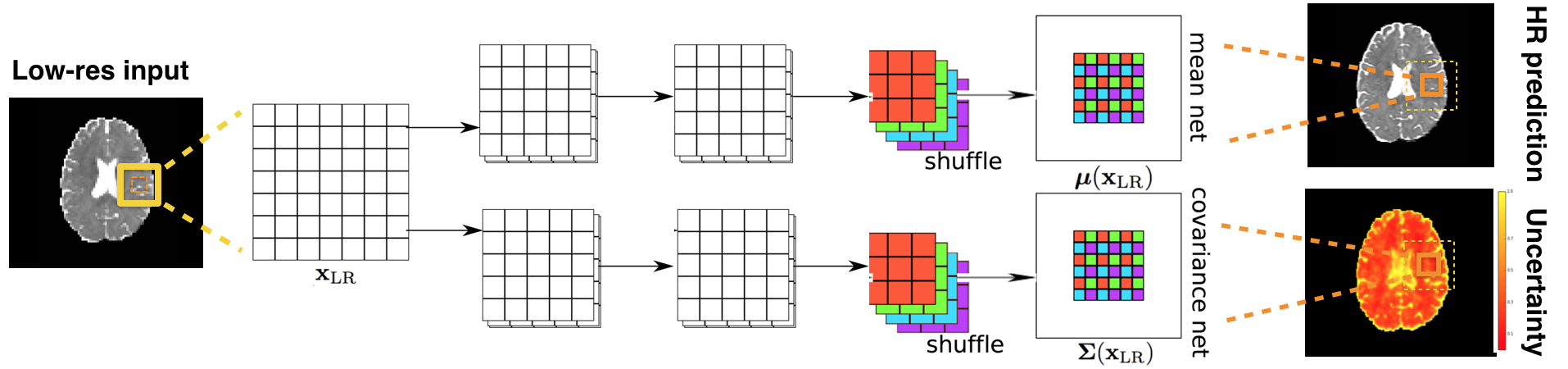}
	\centering	
	\caption{2D illustration of the proposed dual-path architecture which estimates the mean and diagonal covariance of the Gaussian conditional distributions as functions of the input low-resolution subvolume $\mathbf{x}$.  The ``mean network'' $\mu(\cdot)$ at the top generates the high-resolution prediction, while the ``covariance network''  $\Sigma(\cdot)$ at the bottom estimates the corresponding covariance matrix at the selected location in the volume. The diagonal entries of the covariance are used to quantify the intrinsic uncertainty. The parameters of both networks are learned by minimising the common loss function (eq.~\eqref{eq:loss_hetero}).} 
	\label{fig:hetero}
	%\vspace{-7pt}
\end{figure}

\textit{Intrinsic uncertainty} quantifies the inherent ambiguity of the underlying problem that is irreducible with data as illustrated in Fig.~\ref{fig:uncertainty_types}(i). Here we capture intrinsic uncertainty by estimating the variance of the target conditional distribution $p(\mathbf{y}|\mathbf{x}, \theta)$. In medical images, intrinsic uncertainty is often spatially and channel-wise varying. For example, super-resolution could be fundamentally harder on some anatomical structures than others due to signal variability as shown in \cite{tanno2016bayesian}. It may also be the case that some channels of the image volume might contain more complex, non-linear and noisy signals than other channels e.g. higher order terms in diffusion signal representations. To capture such potential variation of intrinsic uncertainty, we model $p(\mathbf{y}|\mathbf{x}, \theta)$ as a Gaussian distribution with input-dependent varying variance: 
\begin{equation}\label{eq:hetero_pdf}
p(\mathbf{y}|\mathbf{x},\mathbf{\theta}_1, \mathbf{\theta}_2) = \mathcal{N}\big{(}\mathbf{y}; \mu(\mathbf{x};\mathbf{\theta}_1),\Sigma(\mathbf{x};\theta_2)\big{)} = \frac{\text{exp} \Big{(} \big{(}\mathbf{y}-\mu(\mathbf{x};\theta_1)\big{)}^\text{T}\Sigma^{-1}(\mathbf{x};\theta_2)\big{(}\mathbf{y}-\mu(\mathbf{x};\theta_1)\big{)}\Big{)}}{\sqrt{(2\pi)^{k} \text{det }\Sigma( \mathbf{x};\theta_2)} }
\end{equation}
 where the mean $\mu(\mathbf{x};\theta_1)$ and the covariance $\Sigma(\mathbf{x};\theta_2)$ are functions of input $\mathbf{x}$ and modelled by two separate 3D-ESPCNs (as shown in 
 Fig.~\ref{fig:hetero}), which we refer to as ``mean network'' and ``covariance network'', and are parametrised by $\theta_1$ and $\theta_2$, respectively. We note that the input patch $\mathbf{x}$ varies spatially, which makes the estimated variance spatially varying and different for respective channels. Fig.~\ref{fig:hetero} shows a 2D illustration of our 3D architecture. For each low-resolution input patch $\mathbf{x}$, we use the output of the mean network $\mu(\mathbf{x};\theta_1)$ at the top as the final estimate of the high-resolution ground truth $\mathbf{y}$ whilst the diagonal elements of the covariance $\Sigma(\mathbf{x};\theta_2)$ quantify the corresponding intrinsic uncertainty over individual components in $\mu(\mathbf{x};\theta_1)$ and over different channels. Lastly, we note that this is a specifc instance of a broad class of models, called \textit{heteroscedastic noise models} \cite{rao1970estimation,nix1994estimating} where the variance  is a function of the value of the input. In contrast, the baseline 3D-ESPCN can be viewed as an example of \textit{homoscedastic noise models} with $\mathbf{y} = \mu_{\theta}(\mathbf{x}) + \sigma \epsilon$, $\epsilon \sim \mathcal{N}(0, I)$ with constant variance $ \sigma^2$ across all spatial locations and image channels, which is highly unrealistic in most medical images.

We jointly optimise the parameters $\mathbf{\theta} = \{\mathbf{\theta}_1, \mathbf{\theta}_2\}$ of the mean network and the covariance network by minimising the negative loglikelihood (NLL): 
\begin{align}
\mathcal{L}_{\theta}(\mathcal{D}) 
&= \sum_{(\mathbf{x}_i,\mathbf{y}_i)\in\mathcal{D}}- \text{log } p(\mathbf{y}|\mathbf{x},\mathbf{\theta}_1, \mathbf{\theta}_2) \\
&= \sum_{(\mathbf{x}_i,\mathbf{y}_i)\in\mathcal{D}}- \text{log } \mathcal{N}\big{(}\mathbf{y}_i; \mu(\mathbf{x}_i;\mathbf{\theta}_1),\Sigma(\mathbf{x}_i;\mathbf{\theta}_2)\big{)} \\
&= \mathcal{M}_{\theta}(\mathcal{D}) + \mathcal{H}_{\theta}(\mathcal{D}) + c \label{eq:loss_hetero}
\end{align}
where $c$ is a constant and the remaining terms are given by
\begin{equation*}
\mathcal{M}_{\theta}(\mathcal{D}) = \frac{1}{N}\sum_{i=1}^{N}\big{(}\mathbf{y}_i-\mu(\mathbf{x}_i;\theta_1)\big{)}^\text{T}\Sigma^{-1}(\mathbf{x}_i;\theta_2)\big{(}\mathbf{y}_i-\mu(\mathbf{x}_i;\theta_1)\big{)}, \quad
\mathcal{H}_{\theta}(\mathcal{D}) = \frac{1}{N}\sum_{i=1}^{N}\text{log det }\Sigma( \mathbf{x}_i;\theta_2).
\end{equation*}
Here $\mathcal{M}_{\theta}(\mathcal{D})$ denotes the mean squared Mahalanobis distance with respect to the predictive distribution $p(\mathbf{y}|\mathbf{x},\mathbf{\theta})$. For simplicity, in this work we assume diagonality of the covariance matrix $\Sigma(\mathbf{x};\theta_2)$. This means that the Mahalanobis distance term $\mathcal{M}_{\theta}(\mathcal{D})$ equates to the sum of MSEs across all pixels and channels in the output, weighted by the inverse of the corresponding variance (estimated intrinsic uncertainty)\footnote{In the case of full covariance, $\mathcal{M}_{\theta}(\mathcal{D})$ becomes the MSE in the basis of principle components, weighted by the corresponding eigenvalues.}. This term naturally encourages assigning low uncertainty to regions with higher MSEs, robustifying the training to noisy labels and outliers. On other other hand, $\mathcal{H}_{\theta}(\mathcal{D})$ represents the mean differential entropy and discourages the spread of $\Sigma_{\theta_2}(\mathbf{x})$ from growing too large. We note that the covariance network is used to modulate the training of the mean network and quantify intrinsic uncertainty during inference while only the mean network generates the final prediction, requiring a single 3D-ESPCN to perform super-resolution.

\subsection{Parameter Uncertainty and Variational Dropout} 
\textit{Parameter uncertainty} signifies the ambiguity in selecting the parameters of the model that best describes the training data as illustrated in Fig.~\ref{fig:uncertainty_types}.(ii). The limitation of the previously introduced 3D-ESPCN baseline (Sec.~\ref{sec:baseline}) and its heteroscedastic extension (Sec.~\ref{sec:hetero}) is their reliance on a single estimate of network parameters. In many medical imaging problems, the amount of training data is modest; in such cases, this point estimate approach increases the risk of overfitting \cite{gal2015dropout}. 

We combat this problem with a Bayesian approach. Specifically, instead of resorting to a single network of fixed parameters, we consider the (posterior) distribution over all the possible settings of network parameters given training data $p(\theta|\mathcal{D})$.  This probability density encapsulates the parameter uncertainty, with its spread of mass describing the ambiguity in selecting most appropriate models to explain the training data $\mathcal{D}$. However, in practice, the posterior $p(\theta|\mathcal{D})$ is intractable due to the difficulty in computing the normalisation constant. We, therefore, propose to approximate $p(\theta|\mathcal{D})$ with a simpler distribution $q_{\phi}(\theta)$ \cite {blei2017variational}. Specifically, we adapt a technique called \textit{variational dropout} \cite{kingma2015variational} to convolution operations from its original version introduced for feedforward NNs. 

Binary dropout \cite{srivastava2014dropout} is a popular choice of method for approximating posterior distributions \cite{gal2015dropout} with demonstrated utility in medical imaging applications \cite{worrall2016automated,yang2016fast,leibig2017leveraging,roy2019bayesian,eaton2018towards,nair2018exploring,bragman2018uncertainty}. However, typically hyper-parameters (dropout rates) need to be pre-set before the training, requiring inefficient cross-validation and thus substantially constraining the flexibility of approximate distribution family $q_{\phi}(\cdot)$ (often a fixed dropout rate per layer). This limitation motivates us to use variational dropout \cite{kingma2015variational}  that extends such approach with a way to learn the dropout rate from data for every single weight in the network and theoretically enables a more effective approximation of the posterior distribution. Another established class of methods is stochastic gradient Markov chain Monte Carlo (SG-MCMC) method \cite{neal1993bayesian,welling2011bayesian,chen2014stochastic,ma2015complete}. However, in this work, we do not not consider SG-MCMC methods because they remain, although unbiased, computationally inefficient due to the requirement of evaluating an ensemble of models for posterior computation, and are slow to converge for high-dimensional problems. 

Variational dropout \cite{kingma2015variational} employs a form of variational inference to approximate the posterior $p(\theta|\mathcal{D})$ by a member of tractable family of distributions $q_{\phi}(\theta) = \prod_{ij} \mathcal{N}(\theta_{ij}; \eta_{ij}, \alpha_{ij}\eta_{ij}^2)$ parametrised by $\phi=\{\eta_{ij}, \alpha_{ij}\}_{ij}$, such that Kullback-Leibler (KL) divergence $\text{KL} \big{(}q_{\phi}(\theta)||p(\theta|\mathcal{D})\big{)}$ is minimised. Here, $\theta_{ij}$ denotes an individual element in the convolution filters of CNNs as a random variable with parameters $\alpha_{ij}$(dropout rate) and $\eta_{ij}$ (mean), and the posterior over the set of all weights is effectively approximated with a product of univariate Gaussian distributions. In practice, introducing a prior $p(\theta)$ and applying Bayes' rule allow us to rewrite the minimization of the KL divergence as maximization of the quantity known as the evidence lower bound (ELBO) \cite{blei2017variational}. Here during training, we learn the variational parameters $\phi=\{\eta_{ij}, \alpha_{ij}\}_{ij}$ by minimizing the negative ELBO (to be consistent with the NLL cost function in eq.(3)): 
 
 \begin{equation}
\mathcal{L}_{\phi} (\mathcal{D}) =  \sum_{(\mathbf{x}_i,\mathbf{y}_i)\in\mathcal{D}} \Big{(}  \mathbb{E}_{q_{\phi}(\theta)}[- \text{log }p(\mathbf{y}_i|\mathbf{x}_i,\theta)]   +  \text{KL} (q_{\phi}(\theta)||p(\theta))\Big{)}
 \end{equation}
An accurate approximation for the KL term for log-uniform prior $p(\theta)$ is proposed in \cite{molchanov2017variational}, which is employed here. On the other hand, the first term (referred to as the reconstruction term) cannot be computed exactly, thus we employ the following MC approximation by sampling $S$ samples of network parameters from the posterior:
\begin{equation}\label{eq:reconstruction}
\mathbb{E}_{q_{\phi}(\theta)}[- \text{log }p(\mathbf{y}|\mathbf{x},\theta)] \approx \frac{1}{S}\sum_{s=1}^{S} - \text{log }p(\mathbf{y}|\mathbf{x},\theta^{(s)}), \quad \theta^{(s)} \sim q_{\phi}(\theta)
\end{equation}
Adapting the local reparametrisation trick presented in \cite{kingma2015variational} to a convolution operation, we derive the implementation of posterior sampling $\theta^{(s)} \sim q_{\phi}(\theta)$ such that the variance of gradients over each mini-batch is low \footnote{See the proof for feedforward networks given in  \cite{kingma2015variational} which generalises to convolutions}. In practice, this amounts to replacing each standard convolution kernel with a ``Bayesian'' convolution, which proceeds as follows. Firstly, we define two separate convolution kernels: $\mathbf{\eta} \in \mathbb{R}^{c\times k^2}$ (``mean'' kernels) and $\alpha \odot \eta^{2} \in \mathbb{R}^{c\times k^2}$ (``variance'' kernels) where $ \odot $ denotes the element-wise multiplication, $c$ is the number of input channel and $k$ is the kernel width. Input feature maps $F_{\text{in}}$ and its elementwise squared values are convolved by respective kernels to compute the ``mean'' and ``variance'' of the output feature maps $\mu_{Y}\delequal F_{\text{in}} \star \mathbf{\eta}$ and $\sigma^{2}_{Y} \delequal F_{\text{in}}^2\star (\alpha \odot \eta^{2})$. Lastly, the final output feature maps $F_{\text{out}}$ are computed by drawing a sample from $\mathcal{N}(\mu_{Y}, \sigma^{2}_{Y})$ i.e. computing the following quantity: 
\begin{equation}\label{eq:noise_injection}
F_{\text{out}} \delequal \mu_{Y} +  \sigma_{Y} \odot \mathbf{\epsilon}, \quad \mathbf{\epsilon} \sim \mathcal{N}(0,I).
\end{equation}
Every forward pass (i.e. computation of each $p(\mathbf{y}|\mathbf{x},\theta^{(s)})$) with variational dropout is thus performed via a sequence of Bayesian convolutions.  Since the injected Gaussian noise $\epsilon$ is independent of the variational parameters $\phi=\{\eta_{ij}, \alpha_{ij}\}_{ij}$, the approximate reconstruction term in eq.~\ref{eq:reconstruction} is differentiable with respect to them \cite{kingma2013auto}.

\subsection{Joint Modelling of Intrinsic and Parameter Uncertainty}
We now describe how to combine the methods for modelling intrinsic and parameter uncertainty. Operationally, we take the dual architecture (Fig.~\ref{fig:hetero}) used to model intrinsic uncertainty, and apply variational dropout to every convolution layer in it. The intrinsic uncertainty is modelled in the heteroscedastic Gaussian model $p(\mathbf{y}|\mathbf{x}, \theta_1, \theta_2) = \mathcal{N}\big{(}\mathbf{y}; \mu(\mathbf{x};\theta_1),\Sigma(\mathbf{x};\theta_2)\big{)}$ while the parameter uncertainty is captured in the approximate posterior $q_{\phi}(\theta_1, \theta_2) \approx p(\theta_1, \theta_2|\mathcal{D}) $ obtained from variational dropout. 

 At test time, for each low-resolution input subvolume $\mathbf{x}$, we would like to compute the predictive distribution $p(\mathbf{y}|\mathbf{x}, \mathcal{D})$ over the high-resolution output $\mathbf{y}$. We approximate this quantity by $q_\phi^*(\mathbf{y}|\mathbf{x})$ by taking the ``average'' of all possible network predictions $p(\mathbf{y}|\mathbf{x}, \theta)=\mathcal{N}\big{(}\mathbf{y}; \mu(\mathbf{x};\theta_1),\Sigma(\mathbf{x};\theta_2)\big{)} $ from all settings of the parameters $\theta_1, \theta_2$, weighted by the associated approximate posterior distribution $q_{\phi}(\theta_1, \theta_2)$. More formally, we need to compute the integral below:
\begin{align} 
q_\phi^*(\mathbf{y}|\mathbf{x}) &\delequal  \int \underbrace{\mathcal{N}(\mathbf{y}; \mu(\mathbf{x};\theta_1),\Sigma(\mathbf{x};\theta_2))) }_{\text{Network prediction}}\cdot \underbrace{q_{\phi}(\theta_1, \theta_2)}_{\text{Approx. posterior}}  \mathrm{d}\theta_1  \mathrm{d}\theta_2  \label{eq:full_distribution} \\
& \approx \int p(\mathbf{y}|\mathbf{x}, \theta_1, \theta_2)\cdot p(\theta_1, \theta_2|\mathcal{D})  \mathrm{d}\theta_1 \mathrm{d}\theta_2 = p(\mathbf{y}|\mathbf{x}, \mathcal{D})  
\end{align}
where the last line represents the true predictive distribution $p(\mathbf{y}|\mathbf{x}, \mathcal{D})$ which is estimated by our model $q_\phi^*(\mathbf{y}|\mathbf{x})$. However, in practice, the integral $q_\phi^*(\mathbf{y}|\mathbf{x})$ cannot be evaluated in closed form because the likelihood $\mathcal{N}\big{(}\mathbf{y}; \mu(\mathbf{x};\theta_1),\Sigma(\mathbf{x};\theta_2)\big{)}$ is a highly non-linear function of input $\mathbf{x}$ as given in eq.~\ref{eq:hetero_pdf}. At test time, we therefore estimate, for each input $\mathbf{x}$, the mean and covariance of the approximate predictive distribution $q_\phi^*(\mathbf{y}|\mathbf{x})$ with the unbiased Monte Carlo estimators: 
\vspace{-3mm}
\begin{align}
\hat{\mu}_{\mathbf{y}|\mathbf{x}}&\delequal \frac{1}{T}\sum_{t=1}^T\mu(\mathbf{x};\theta_{1}^{t}) \label{eq:est_mean} \xrightarrow[T\rightarrow\infty]{}\mathbb{E}_{q_\phi^*(\mathbf{y}|\mathbf{x})} [\mathbf{y}]\\
\hat{\Sigma}_{\mathbf{y}|\mathbf{x}} &\delequal \frac{1}{T} \sum_{t=1}^T\Big{(}\Sigma(\mathbf{x};\theta_{2}^{t})+\mu(\mathbf{x};\theta_1^t)\mu(\mathbf{x};\theta_1^t)^{\text{T}}\Big{)}-\hat{\mu}_{\mathbf{y}|\mathbf{x}} \hat{\mu}_{\mathbf{y}|\mathbf{x}}^{\text{T}}\xrightarrow[T\rightarrow\infty]{}\text{cov}_{q_\phi^*(\mathbf{y}|\mathbf{x})} [\mathbf{y}, \mathbf{y}]
\label{eq:est_cov}
\end{align}
where $\{(\theta^t_1,\theta^t_2)\}_{t=1}^T$ are samples of the network parameters (i.e. convolution kernels) drawn from the approximate posterior $ q_{\phi}(\theta_1, \theta_{2})$. In the other words, the inference performs $T$ stochastic forward passes at test time by injecting noise into features according to eq.~\ref{eq:noise_injection}, and amalgamates the corresponding network outputs to compute the sample mean $\hat{\mu}_{\mathbf{y}|\mathbf{x}}$ and sample covariance $\hat{\Sigma}_{\mathbf{y}|\mathbf{x}}$. We use the sample mean $\hat{\mu}_{\mathbf{y}|\mathbf{x}}$  as the final prediction of an high-resolution ouput patch $\mathbf{y}$ and use the diagonal elements of the sample covariance $\hat{\Sigma}_{\mathbf{y}|\mathbf{x}}$ to quantify the corresponding uncertainty, which we refer to as \textit{predictive mean} and \textit{predictive uncertainty}, respectively. 

%When we use the baseline 3D-ESPCN, the first term in the sample variance reduces to $\sigma^2I$, which is equivalent to homogeneous intrinsic noise. 

\begin{figure}
	%\vspace{-20pt}
	\includegraphics[width=\linewidth]{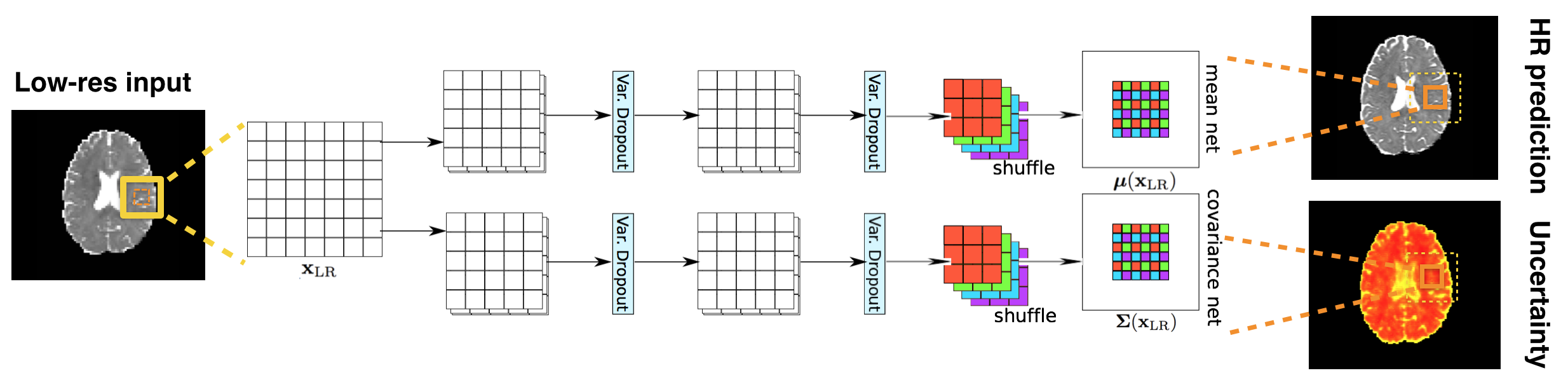}
	\centering	
	\small 
	\vspace{-8mm}
	\caption{\small 2D illustration of a heteroscedastic network with variational dropout. Diagonal covariance is again assumed. The top 3D-ESPCN estimates the mean and the bottom one estimates the covariance matrix of the likelihood. Variational dropout is applied to feature maps after every convolution where Gaussian noise is injected into feature maps $F_{\text{out}} = \mu_{Y} +  \sigma_{Y} \odot \mathbf{\epsilon}$ where $\mathbf{\epsilon} \sim \mathcal{N}(0,I)$ (see eq.~\ref{eq:noise_injection}). } 
	\label{fig:heterovar}
	%\vspace{-7pt}
\end{figure}

\subsection{Uncertainty Decomposition and Propagation} \label{sec:uncertainty_decom}
Predictive uncertainty arises from the combination of two source effects, namely intrinsic and parameter uncertainty, for which we have previously introduced methods for estimation. Lastly, we introduce a method based on variance decomposition for disentangling these effects and quantifying their contributions separately in predictive uncertainty. We consider such decomposition problem in the presence of an arbitrary transformation of the output variable $\mathbf{y}$. 

The users of super-resolution algorithms are often interested in the quantities that are derived from the predicted high-resolution images, rather than the images themselves. For example, quantities such as the principal direction (first eigenvalue of the DT), mean diffusivity (MD) and fractional anisotropy (FA) are typically calculated from diffusion tensor images (DTIs) and used in the downstream analysis. We therefore consider an generic function\footnote{We assume here that the transform $g$ is a measurable function with well-defined expectation and variance.} $g:\mathcal{Y}\rightarrow\mathbb{R}^{m}$ which transforms the high-resolution multi-channel data $\mathbf{y}$ to a quantity of interest e.g. MD and FA maps, and propose a way to propagate the predictive uncertainty over $\mathbf{y}$ to the transformed domain (i.e. compute the variance of $p(g(\mathbf{y})|\mathcal{D}, \mathbf{x})$) and decompose it into the ``intrinsic'' and ``parameter'' components. Specifically, by using the law of total variance \cite{weiss2006course}, we perform the following decomposition:

%\begin{align}
%\mathbb{V}_{p(\mathbf{y}|\mathbf{x},\mathcal{D})} [g(\mathbf{y})]
%& =  %\mathbb{E}_{p(\theta|\mathcal{D})}\big{[}\mathbb{V}_{p(g(\mathbf{y})|\theta,\mathbf{x},\mathcal{D})}[g(\mathbf{y})]-\mathbb{V}_{p(g(\mathbf{y})|\theta,\mathbf{x},\mathcal{D})}[g(\mathbf{y})|\theta]\big{]} + \mathbb{E}_{p(\theta|\mathcal{D})}\big{[}\mathbb{V}_{p(g(\mathbf{y})|\theta,\mathbf{x},\mathcal{D})}[g(\mathbf{y})|\theta]\big{]}\\
%&=\underbrace{\mathbb{V}_{p(\theta|\mathcal{D})}[\mathbb{E}_{p(g(\mathbf{y})|\theta,\mathbf{x},\mathcal{D})}[g(\mathbf{y})|\theta]]}_{\text{propagated parameter uncertainty}} +   \underbrace{\mathbb{E}_{p(\theta|\mathcal{D})}[\mathbb{V}_{p(g(\mathbf{y})|\theta,\mathbf{x},\mathcal{D})}[g(\mathbf{y})|\theta]]}_{\text{propagated intrinsic uncertainty}} \\
%&= \Delta_{m}(g(\mathbf{y})) + \Delta_{i}(g(\mathbf{y}))  
%\label{eq:variance_decomposition}
%\end{align}

\begin{equation}
\mathbb{V}_{p(\mathbf{y}|\mathbf{x},\mathcal{D})} [g(\mathbf{y})] = \Delta_{m}(g(\mathbf{y})) + \Delta_{i}(g(\mathbf{y}))  
\label{eq:variance_decomposition}
\end{equation}
where the respective component terms are given by:
\begin{align}
\Delta_{m}(g(\mathbf{y})) & = \mathbb{E}_{p(\theta|\mathcal{D})}\big{[}\mathbb{V}_{p(g(\mathbf{y})|\theta,\mathbf{x},\mathcal{D})}[g(\mathbf{y})]-\mathbb{V}_{p(g(\mathbf{y})|\theta,\mathbf{x},\mathcal{D})}[g(\mathbf{y})|\theta]\big{]} \\
&=\underbrace{\mathbb{V}_{p(\theta|\mathcal{D})}[\mathbb{E}_{p(g(\mathbf{y})|\theta,\mathbf{x},\mathcal{D})}[g(\mathbf{y})|\theta]]}_{\text{propagated parameter uncertainty}} 
\\
\Delta_{i}(g(\mathbf{y}))  &=   \underbrace{\mathbb{E}_{p(\theta|\mathcal{D})}[\mathbb{V}_{p(g(\mathbf{y})|\theta,\mathbf{x},\mathcal{D})}[g(\mathbf{y})|\theta]]}_{\text{propagated intrinsic uncertainty}} 
\end{align}
We refer to the components $ \Delta_{m}(g(\mathbf{y})) $ and $ \Delta_{i}(g(\mathbf{y})) $ as ``propagated'' parameter and intrinsic uncertainty.  Intuitively, the first term quantifies the difference in variance between the cases where we have variable parameters and fixed parameters. In other words, this quantifies how much predictive uncertainty on the derived quantity arises, on average, from the variability in parameters. The second term on the other hand quantifies the average variance of the model prediction when the parameters are fixed, which signifies the model-independent uncertainty due to data i.e. intrinsic uncertainty.  Assuming that the considered neural network is identifiable\footnote{We note that a neural network is, in general, not identifiable i.e. there exist more than a single set of parameters that capture the same target distribution $p(g(\mathbf{y})|\mathbf{x})$. In such cases, the posterior distribution $p(\theta|\mathcal{D})$ does not collapse to a single Dirac Delta function with infinite amount of observations---it rather converges to a mixture of all sets of network parameters $\Theta$  such that $p(g(\mathbf{y})|\theta^{*},\mathbf{x}) = p(g(\mathbf{y})|\mathbf{x}) \forall \theta^{*} \in \Theta $. However, the expectation $\mathbb{E}_{p(g(\mathbf{y})|\theta,\mathbf{x},\mathcal{D})}[g(\mathbf{y})|\theta]$ is the same for all $\theta \in \Theta$ and thus the propagated parameter uncertainty $ \Delta_{m}(g(\mathbf{y}))$ converges to zero. } and sufficiently complex to capture the underlying data generating process, as the amount of training data increases, the posterior $p(\theta|\mathcal{D})$ tends to a Dirac delta function and thus the first term diminishes to zero while the second term remains. A similar variance decomposition technique was employed in \cite{bowsher2012identifying} to understand how the variation in cell signals of interest (e.g. gene expression) in a bio-chemical network is caused by the fluctuations of other environmental variables (e.g. transcription rate and biological noise). In our case, we employ the variance decomposition technique to separate the effects of network parameters from the intrinsic uncertainty in the prediction of $g(\mathbf{y})$.  

%In statistics \cite{weiss2006course}, the two terms are described as the variance components  ``explained'' and ``unexplained'' by the model parameters.

We first consider a special case where the transform $g$ is an identify map i.e. $g(\mathbf{y})=\mathbf{y}$. Assuming the likelihood is modelled by a Gaussian distribution with heteroscedastic noise i.e. $p(\mathbf{y}|\theta_1, \theta_2, \mathbf{x},\mathcal{D}) = \mathcal{N}(\mathbf{y}; \mu(\mathbf{x};\theta_1),\Sigma(\mathbf{x};\theta_2))$, then we can show that the parameter and intrinsic uncertainty are given by
\begin{equation}
\Delta_{m}(\mathbf{y}) = \mathbb{V}_{p(\theta_{1}|\mathcal{D})}[\mu_{\theta_1}(\mathbf{x})], \quad \Delta_{i}(\mathbf{y})  =  \mathbb{E}_{p(\theta_{2}|\mathcal{D})}[\Sigma_{\theta_2}(\mathbf{x})]
\end{equation}
which can be approximated by the components of the MC variance estimator in eq.~\eqref{eq:est_cov} :
\begin{align}\label{eq:sampler_mean}
\widehat{\Delta}_{m}(\mathbf{y})  &=\frac{1}{T} \sum_{t=1}^T\mu(\mathbf{x};\theta_1^t)\mu(\mathbf{x};\theta_1^t)^{\text{T}} -\hat{\mu}_{\mathbf{y}|\mathbf{x}} \hat{\mu}_{\mathbf{y}|\mathbf{x}}^{\text{T}}\\ 
\widehat{\Delta}_{i}(\mathbf{y})  &= \frac{1}{T} \sum_{t=1}^T\Sigma(\mathbf{x};\theta_{2}^{t}) \label{eq:sampler_var}
\end{align}
where $\{(\theta^t_1,\theta^t_2)\}_{t=1}^T$ are drawn from the approximate posterior $q_{\phi}(\theta_1, \theta_{2})$. %If you turn off the variaitional dropout in the covariance network, the intrinsic uncertainty term reduces to $\Delta_{i}(\mathbf{y}) = \Sigma_{\theta_2^*}(\mathbf{x})$ where $\theta_2^*$ denotes the point-estimate of parameters in the covariance network. 

%\textcolor{red}{Food for thought: May be you can generalise this argument for transforms $g$ of specific types e.g. low-variance estimators similar to eq. \eqref{eq:sampler_mean} and eq. \eqref{eq:sampler_var} when $g$ additively or multiplicatively separates model parameters $\theta$ between heteroschedastic noise and posterior distribution? e.g. MD is a good example }.

More generally, when the transform $g$ is complicated, MC sampling provides an alternative implementation. Given samples of model parameters $\{\theta_{t}\}_{t=1}^{T} \sim q(\theta|\mathcal{D})$ and $\{g^{t}_{j}\}_{j=1}^J \sim p(g(\mathbf{y})|\theta_{t},\mathbf{x},\mathcal{D})$ for $t=1, ...,T$,  we estimate both the progapated parameter and intrinsic uncertainty as follows:

% Before bessel's correction
%\begin{align}
%\widehat{\Delta}_{m}(g(\mathbf{y}))  &\delequal \frac{1}{T}\sum_{t}(\hat{\mu}^{t})^2  - \Big{(}\frac{1}{JT}\sum_{j,t}(g^{t}_{j}) \Big{)}^2 \label{eq:mc_model_uncertainty} \\ 
%\widehat{\Delta}_{i}(g(\mathbf{y}))  &\delequal  \frac{1}{JT}\sum_{j,t}(g^{t}_{j})^2 - \frac{1}{T}\sum_{t}(\hat{\mu}^{t})^2 \label{eq:mc_intrinsic_uncertainty} \\
%\hat{\mu}^{t} & = \frac{1}{J}\sum_{j}g^{t}_{j}
%\end{align}

\begin{align}
\widehat{\Delta}_{m}(g(\mathbf{y}))  &\delequal \frac{1}{T}\sum_{t}(\hat{\mu}^{t})^2  - \Big{(}\frac{1}{(J-1)T}\sum_{j,t}(g^{t}_{j}) \Big{)}^2 \label{eq:mc_model_uncertainty} \\ 
\widehat{\Delta}_{i}(g(\mathbf{y}))  &\delequal  \frac{1}{(J-1)T}\sum_{j,t}(g^{t}_{j})^2 - \frac{1}{T}\sum_{t}(\hat{\mu}^{t})^2 \label{eq:mc_intrinsic_uncertainty} \\
\hat{\mu}^{t} & = \frac{1}{J}\sum_{j}g^{t}_{j}. 
\end{align}
These estimators are, although unbiased, higher in variance than the case where $g$ is the identity (eq.~\eqref{eq:sampler_mean} and eq.~\eqref{eq:sampler_var}), due to two sources of sampling, thus requiring more samples for reliable estimation of respective uncertainty components. 

%\textcolor{red}{These estimators are biased at the moment - need to check this properly but should be fine just to replace $J$ with $J-1$. Also, we should mention that these estimators are high in variance due to two sources of sampling, and requires more samples than eq. \eqref{eq:sampler_mean} and eq. \eqref{eq:sampler_var}}.

%\subsection{Multiscale Model with Cropping and Recursive Connections}
%\textcolor{red}{This section is only needed if multi-scale architectures are included in the journal paper.}

% ---------------- Experiments ----------------------------
\section{Experiments and Results}

In this section, we evaluate the proposed uncertainty modelling techniques for super-resolution of diffusion MR images. We first compare quantitatively the reconstruction performance of our probabilistic CNN models against the relevant baselines in two different types of diffusion signal representations. Secondly, we study the real-world utility of the technique in downstream tractography applications. Thirdly, we evaluate the value of predictive uncertainty as a realiability metric of output images on multiple datasets of both healthy subjects and those with unseen pathological structures such as brain tumour (Glioma) and multiple sclerosis (MS). 

%Lastly, we decompose the measure of predictive uncertainty into the intrinsic and parameter components, and correlate them to the prediction accuracy on a HCP healthy subject with benign abnormalities.  \textcolor{red}{come back to this after modifying the section \ref{sec:unseen_abnormality}}

% We 

%This section focuses on the task of diffusion MRI super-resolution and evaluates the utility of the proposed uncertainty modelling techniques in terms of reconstruction accuracy and risk management of preditive failures. The accuracy of the baseline CNN architecture (3D-ESPCN) and its probabilistic extensions are compared against standard interpolation techniques and random-forest implementations of IQT \cite{alexander2017image,tanno2016bayesian}. In particular, we demonstrate on two different types of diffusion MR images, namely diffusion tensor images (DTIs) and Mean Apparent Propagator (MAP)-MRI \cite{ozarslan2013mean}. We also study the effects of uncertainty modeling on generalisability by testing on the Lifespan dataset, which differs from the training data in acquisition protocl and population range. We additionally test the utility of improved model accuracy in downstream tractography applications. We then study how the value of predictive uncertainty provides an indicator of predictive accuracy on images of both healthy subjects and brain tumour patients. Lastly, we decompose the measure of predictive uncertainty into the intrinsic and parameter components, and correlated them to the model performance on the DTI of a HCP healthy subject with benign abnormalities.

\subsection{Datasets} \label{sec:dataset}
We make use of the following four diffusion MRI datasets to evaluate different benefits of the proposed technique: 

\begin{itemize}
	\item \textbf{Human Connectome Project dataset: } we use the diffusion MRI data from the WU-Minn HCP (release Q3) \cite{van2013wu} as the source of the training datasets. The dataset enjoys very high image resolution, signal levels and coverage of the measurement space, enabled by the combination of custom imaging, reconstruction innovations and a lengthy acquisition protocol \cite{sotiropoulos2013advances}. Each subject's data set contains $288$ diffusion weighted images (DWIs) of voxel size $1.25^{3} \text{ mm}^{3}$ of which $18$ have nominal $b=0$ and the three high-angular-resolution-diffusion-imaging (HARDI) shells of 90 directions have nominal b-values of $1000, 2000$, and $3000 \text{ smm}^{-2}$ (see \cite{sotiropoulos2013advances} for the full acquisition details). The data are preprocessed by correcting distortions including susceptibility-induced, eddy currents and motion as outlined in \cite{glasser2013minimal}.
	
	\item \textbf{Lifespan dataset:} this dataset (available online at \url{http://lifespan.humanconnectome.org} contains $26$ subjects of much wider age range ($8-75$ years) than the main HCP cohorts ($22-36$ years), and is acquired with a shortened version of the main HCP protocol with lower resolution ($1.5$ mm isotropic voxels) and only two HARDI shells, with $b = 1000$ and $2500 \text{ smm}^{-2}$. However, we also note that the protocol still leverages the special features of the HCP scanners, providing images of substantially better quality than standard sequences. We utilise this out-of-training-distribution dataset  to assess the robustness of our techniques to domain shifts. 
	 
	\item \textbf{Prisma dataset:} two healthy male adults (29 and 33 years old respectively) were scanned twice at different image resolutions using the clinical 3T Siemens Prisma scanner in FMRIB, Oxford. Both datasets contain diffusion MRI data with 21 $b=0$ images and three 90-direction HARDI shells, b-values of 1000, 2000, and $3000 \text{ smm}^{-2}$, each for two resolutions, 2.50 mm and 1.35 mm isotropic voxels  (see \cite{alexander2017image} for full acquisition details). In addition, each of these datasets also includes a standard 3D T1-weighted MPRAGE (1 mm isotropic resolution). The Prisma scanner is less powerful than the bespoke HCP scanner and cannot achieve sufficient signal at $1.25$ mm resolution, but the $1.35$ mm data provides a pseudo ground-truth for IQT resolution enhancement of the 2.5 mm data. 
	
	\item  \textbf{Pathology dataset:} we use two separate datasets which consist of images of brain tumour (Glioma) \cite{figini2018prediction} and multiple sclerosis (MS) patients, respectively. The data of each wubject with glioma contains DWIs with $b=700 \text{ s/mm}^2$ while the measurement of each MS patient is of $b=1200 \text{ s/mm}^2$. Both datasets have isotropic voxel size $2^3 \text{ mm}^3$, which is closer to the image resolution of commonplace clinical scanners. We use these datasets to assess the behaviour of predictive uncertainty on images with pathological features that are not represented in the training data set.
\end{itemize}

In all the experiments, super-resolution are performed on diffusion parameter maps derived from the DWIs in the above datasets. In particular, we consider two diffusion MRI models, namely the diffusion tensor (DT) model \cite{basser1994mr} and Mean Apparent Propagator (MAP) MRI \cite{ozarslan2013mean}, where the former is the simplest and most standard diffusion parameter map, and the latter is a high-order generalisation of the former with the capacity to characterise signals from more complex tissue structures (e.g. fibre crossing regions), a requirement for successful tractography applications. We compute both of these diffusion parameter maps using the implementation from \cite{alexander2017image}, which is available at \url{https://github.com/ucl-mig/iqt}. 

We fit the DT model to the combination of $b=0$ images and $b=1000  \text{ s/mm}^2$ HARDI shell for the HCP and Lifespan datasets, and $b=700  \text{ s/mm}^2$ shell for the brain tumour dataset. In all cases, weighted linear least squares are employed for the fitting, taking into account the spatially varying b-values and gradient directions in the HCP dataset. On the other hand, in the case of MAP-MRI, $22$ coefficients of basis functions up to order $4$ are estimated via (unweighted) least squares to all three shells of the HCP, Lifespan and Prisma datasets. As noted in \cite{alexander2017image}, the choice of scale parameters (see \cite{ozarslan2013mean}) $\mu_{x} = \mu_{y} = \mu_{z} = 1.2\times 10^{-3}$ mm empirically minimises the fitting error in the HCP dataset, and is used for all datasets. 

Training datasets in all experiments are constructed by artificially downsampling very high-resolution images in the HCP dataset. In particular, we employ the following downsampling procedure: (i) the raw DWIs of selected subjects are blurred by applying the mean filter of size $r\times r \times r$ independently over channels with $r$ denoting the upsampling rate; (ii)  the DT or MAP parameters are computed for every voxel; (iii) the spatial resolution of the resultant parameter maps are reduced by taking every $r$ pixels. A coupled library of low-resolution and high-resolution patches is then constructed by associating each patch in the downsampled DTI/MAP-MRI with the corresponding patch in the ground truth DTI or MAP-MRI. In this case, we ensure the low-resolution patch to be centrally and entirely contained within the corresponding high-resolution patch (as illustrated by the yellow and orange squares in Fig.~\ref{fig:ESPCN}). We then randomly select a pre-set number of patches from each subject in the training pool to create a training dataset as detailed in Table~\ref{tab:train_data}. In addition to the $8$ subjects used in the prior work \cite{alexander2014image,tanno2016bayesian,tanno2017bayesian}, we randomly select additional $8$ subjects from the HCP cohort and include them in the training subject pool. Patches are standardized channel-wise by subtracting the mean of foreground pixel intensities of the corresponding subject and dividing by its standard deviation. Moreover, since MAP-MRI datasets contain outliers due to model fitting, in large enough quantity to influence the training of the baseline 3D-ESPCN model, we remove them by clipping the voxel intensity values of the respective $22$ channels separately at $0.1\%$ and $99.9\%$ percentiles computed over all the foreground voxels in the whole training dataset. 

\begin{table}
	\caption{Details of training data for two diffusion MR signal representations, DTIs and MAP-MRIs. The first two columns from the right denote the size of the input $\mathbf{x}$ and output patches $\mathbf{y}$ of dimension $[\text{width}, \text{height}, \text{depth}, \text{channels}]$ while the third and the fourth columns show the number of patch pairs $(\mathbf{x}, \mathbf{y})$ extracted from each subject, and the total number of training subjects used, respectively. }
	\center
	\small
	\begin{tabular}{lclclclc|cl}
		\hline
		Data &  Size of input $\mathbf{x}$ &Size of output $\mathbf{y}$ &  No. pairs $(\mathbf{x}, \mathbf{y})$ per subject& No. subjects   \\ 
		\hline	
		DTIs &  $11\times 11 \times 11 \times 6$  & $14\times 14 \times 14 \times 6$  & $8000$ & $16$   \\
		MAP-MRIs  & $21\times 21 \times 21 \times 22$  & $14\times 14 \times 14 \times 22$  & $4000$ & $16$  \\
		
		\hline
		\label{tab:train_data}
	\end{tabular}%	
	\vspace{-5mm}

\end{table}

\subsection{Network Architectures and Training}
For the training of all CNN models, we minimised the associated loss function using Adam \cite{kingma2014adam} for $200$ epochs with initial learning rate of $10^{-3}$ and $\beta= [ 0.9, 0.999]$, with minibatches of size $12$. We hold out 50\% of training patch pairs as a validation set. The best performing model was selected based on the mean-squared-error (MSE) on the validation set. 

For the super-resolution of DTIs, as in \cite{shi2016real}, we use a minimal architecture for the baseline 3D-ESPCN, consisting of three $3D$ convolutional layers with filters $(3^3,50)\to(1^3,100)\to(3^3,6r^3)$ where $r$ is upsampling rate and $6$ is the number of channels in DTIs. As illustrated in Fig.~\ref{fig:ESPCN}, the dimensions of convolution filters are chosen, so each $5^3\cdot 6$ low-resolution receptive field patch maps to a $r^3 \cdot 6$ high-resolution patch, which mirrors competing random forest based methods \cite{alexander2014image,tanno2016bayesian} for a fair comparison. On the other hand, for MAP-MRI, which is a more complex image modality with $21$ channels, we employ a deeper model with 6 convolution layers $(5^3, 256)\to(3^3,256)\to(3^3,128)\to(3^3,128)\to(3^3,64)\to(3^3,21r^3)$ prior to the shuffling operation, which expands the receptive field on each $r^3\cdot21$ high-resolution patch to $15^3\cdot 21$ input low-resolution patch. Every convolution layer is followed by a ReLU non-linearity except the last one in the architecture, and batch-normalization \cite{ioffe2015batch} is additionally employed for MAP-MRI super-resolution between convolution layer and ReLU non-linearity. 

The mean and variance networks in the heteroscedastic noise model introduced in Sec.~\ref{sec:hetero} are implemented as two separate baseline 3D-ESPCNs of the architectures, specified above for DTIs and MAP-MRIs. Positivity of the variance is enforced by passing the output through a softplus function $f(x)=\text{ln}(1+e^{x})$ as in \cite{lakshminarayanan2017simple}. 

For variational dropout, we considered two flavours: Var.(I) optimises per-weight dropout rates, and Var.(II) optimises per-filter dropout rates. More formally, the ``drop-out rate" $\alpha_{ij}$ in the approximate posterior $q_\phi(\theta_{ij}) = \mathcal{N}(\theta_{ij}; \eta_{ij}, \alpha_{ij}\eta_{ij}^2)$ is different for every element in each convolution kernel in the former while the latter has common $\alpha_{ij}$ shared across each kernel. In preliminary analysis, we found that the number of samples per data point for estimating reconstruction term (eq.~\ref{eq:reconstruction}) can be set to $S=1$ so long as the batch size is sensibly large ($M=12$). 

We also note the default training with binary and Gaussian dropout also employs $S=1$ \cite {srivastava2014dropout} along with other MC variational inference methods for neural networks such as \cite{kingma2013auto,kingma2015variational,gal2017concrete}. Variational dropout is applied to both the baseline and heteroscedastic models without changing the architectures. For both binary and Gaussian dropout modes, we incorporate the dropout operations of fixed rate $p$ in every convolution layer of the baseline 3D-ESPCN architecture. 

All models are trained on simulated datasets generated from 16 HCP subjects as detailed in Sec.~\ref{sec:dataset}. We also retrained the random forest models employed in \cite{tanno2016bayesian,alexander2017image} on equivalent datasets. It takes under $60/360$ mins to train a single network on DTI/MAP-MRI data on a single TITAN X GPU. All models are implemented in the TensorFlow framework \cite{abadi2016tensorflow} and the codes will be released at \url{https://github.com/rtanno21609/UncertaintyNeuroimageEnhancement}.

\subsection{Quantitative Evaluation of Super-resolution Performance}
\begin{figure}[t]
	%\vspace{- 15pt}
	\centering
	\includegraphics[width=2.75cm]{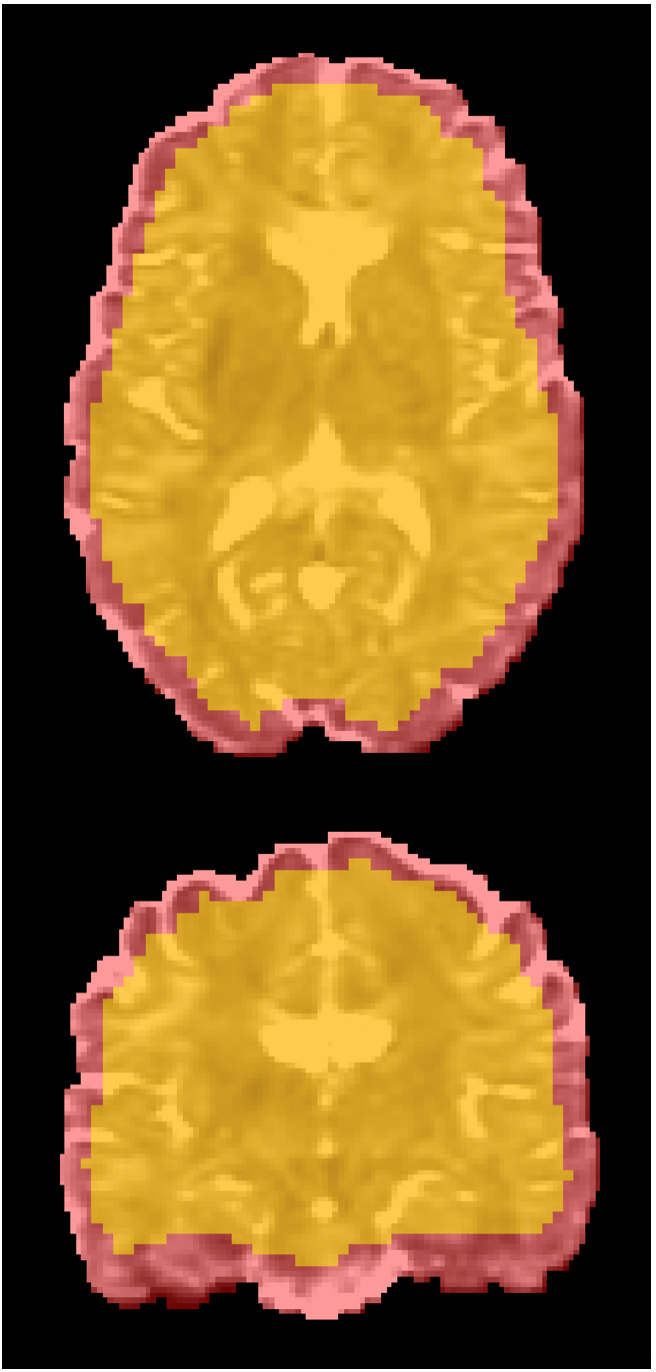}
	\small
	\caption{Visualisation of ``interior'' (yellow) and ``exterior'' regions (red). The interior region consists of a set of patches contained entirely within the brain while the exterior region consists of partial patches that contain mixtures of brain and background voxels} 
	\label{fig:mask}
	%\vspace{- 15pt}
\end{figure}

We evaluate the prediction performance of our models for super-resolution of DTI and MAP-MRI on two datasets---HCP and Lifespan as detailed in Sec.~\ref{sec:dataset}. The first dataset contains 16 unseen subjects from the same HCP cohort used for training, while the second one consists of $10$ subjects from the HCP Lifespan dataset. The latter tests generalisability, as they are acquired with a different protocol at lower resolution (1.5 mm isotropic), and contain subjects of a different age range (45-75 years) to the original HCP data (22-36 years). We perform $\times 2$ upsampling in all spatial directions. The reconstruction quality is measured with root-mean-squared-error (RMSE), peak-signal-to-noise-ratio (PSNR) and mean-structural-similarity (MSSIM) \cite{wang2004image} on two separate regions:  i) ``interior''; set of patches contained entirely within the brain mask; ii) ``exterior''; set of patches containing some brain and some background voxels, as shown in Fig.~\ref{fig:mask}. This is because the current state-of-the-art methods based on random forests (RFs) such IQT-RF \cite{alexander2017image} and BIQT-RF \cite{tanno2016bayesian} are only trained on patches from the interior region and requires a separate procedure on the brain boundary. In addition, the estimation problem is quite different in boundary regions, but remains valuable particularly for applications such as tractography where seed or target regions are often in the cortical surface of the brain. We only present the RMSE results, but the derived conclusions remain the same for the other two metrics. Aside from the interpolation techniques, for each method an ensemble of $10$ models are trained on different trainings set (generated by randomly extracting patch pairs from the common $16$ HCP training subjects) and for each model, the average error metric over the test subjects are first calculated. The mean and standard deviations of such average errors are computed across the model ensemble and reported in Table~\ref{tab:compare_1} and Table~\ref{tab:compare_2}. 

Table~\ref{tab:compare_1} shows that our baseline achieves $8.5\%/39.8\%$ reduction in RMSE for the super-resolution of DTIs on the HCP dataset on the interior/exterior regions with respect to the best published method, BIQT-RF\cite{tanno2016bayesian}. While the standard deviations are higher, the improvements are more pronounced in MAP-MRI super-resolution, reducing the average RMSEs by $49.6\%$ and $63.5\%$ on the interior and exterior regions. We note that that IQT-RF and BIQT-RF are only trained on interior patches, and super-resolution on boundary patches requires a separate \textit{ad hoc} procedure. Despite including exterior patches in training our model, which complicates the learning task, the baseline CNN out-performs the RF methods on both regions. We see similar improvements in the out-of-distribution Lifespan dataset.

Reconstruction is faster than the RF baselines; the 3D-ESPCN is capable of estimating the whole high-resolution DTI/MAP-MRI under 10/60 seconds on a CPU and $1/10$ second(s) on a GPU. On the other hand, BIQT-RF takes $\sim10$ mins with 8 trees on both DTIs and MAP-MRIs. The fully convolutional architecture of the model enables to process input patches of different size from that of training inputs, and we achieve faster reconstruction by using larger input patches of dimension $25^3\cdot c$ where $c$ is the number of channels. We also note that the reconstruction time of the variational dropout based models increases by a factor of the number of MC samples used at test time, although it is possible, with more memory, to leverage GPU parallelisation by making multiple copies of each input patch and treating them as a mini-batch. On the other hand, the heteroscedastic CNN enjoys the same inference speed of the baseline since only the mean network is used for reconstruction (the covariance network is only employed to quantify the estimated intrinsic uncertainty). 

\begin{table}
	\center
	\footnotesize
	\begin{tabular}{@{}lclclclcl}
		\hline
		Models & HCP (interior) & HCP (exterior)& Life (interior) & Life (exterior)   \\ 
		\hline	
		CSpline-interpolation & $10.069 \pm$ n/a & $31.738\pm$ n/a & $32.483\pm$ n/a & $49.066\pm$ n/a  \\
		$\beta$-Spline interpolation& $9.578\pm$ n/a  & $98.169\pm$ n/a &$33.429\pm$ n/a &$186.049\pm$ n/a   \\
		IQT-RF & $ 6.974 \pm 0.024$  & $23.139\pm0.351$ & $10.038\pm0.019$&$25.166\pm0.328$   \\ 
		BIQT-RF & $6.972 \pm 0.069$  &$23.110\pm0.362$  & $9.926\pm0.055$ &$25.208\pm0.290$ \\
		\hline 
		3D-ESPCN(baseline)                      & $6.212\pm0.017$    & $13.609\pm0.084$  &$8.902\pm0.020$   &$16.389\pm0.114$\\
		+ Binary Dropout ($p=0.1$)           & $6.319 \pm 0.015$  & $13.738\pm0.048$   &$9.093\pm0.024$  &$16.489\pm0.099$ \\
		+ Gaussian Dropout ($p=0.05$)     & $6.463\pm0.034$    & $14.168\pm0.051$  &$9.184\pm0.048$  & $16.653\pm0.092$\\
		
		+ Variational Dropout (I)                  & $6.194\pm0.013$    & \cellcolor{red!15} $\mathbf{13.412\pm 0.041}$   &$8.874\pm0.027$  &  \cellcolor{red!15} $\mathbf{16.147\pm0.051}$ \\
		+ Variational Dropout (II)                 & $6.201\pm0.015$    & \cellcolor{blue!15}  $13.479\pm0.047$   & $8.878\pm0.031$  &\cellcolor{blue!15} $16.230\pm0.075$ \\
		
		+ Hetero.                                        & $6.135\pm0.029 $   & $15.469\pm 0.231$  & $8.885\pm0.041$ & $17.208\pm0.211$ \\
		
		+ Hetero. + Variational Dropout (I)   & \cellcolor{blue!15} $6.121\pm0.015$    &$13.591\pm0.051$   & \cellcolor{red!15} $\mathbf{8.837\pm0.043}$ &$16.261\pm0.053$\\ 
		+ Hetero. + Variational Dropout (II)  & \cellcolor{red!15} $\mathbf{6.116\pm0.013}$   &$13.622\pm0.099$   & \cellcolor{blue!15} $8.861\pm0.031$  &$16.387\pm0.098$\\
% The previous experiment with 8 train subjects
%		3D-ESPCN(baseline)                   & $6.378\pm0.015$   & $13.909\pm0.071$   &  $8.998\pm0.021$   &$16.779\pm0.109$\\
%	   	+ Binary Dropout ($p=0.1$)        & $6.519 \pm 0.015$ & $14.038\pm0.038$    & $9.183\pm0.024$   &$16.890\pm0.097$ \\
%		+ Gaussian Dropout ($p=0.05$) & $6.963\pm0.034$   & $14.568\pm0.068$    & $9.784\pm0.048$   &$17.357\pm0.091$\\
%		+ Variational Dropout (I)& $6.354\pm0.015$ & \cellcolor{red!15} $\mathbf{13.824\pm 0.031}$ &$8.973\pm0.024$&\cellcolor{red!15}$\mathbf{16.633\pm0.053}$ \\
%		+ Variational Dropout (II) & $6.356\pm0.008$ & \cellcolor{blue!15} $13.846\pm0.017$  & $8.982\pm0.024$ &\cellcolor{blue!15}$16.738\pm0.073$ & \\
%		+ Hetero.  &$6.294\pm0.029 $ & $15.569\pm 0.273$  & $8.985\pm0.051$ & $17.716\pm0.277$ \\
%		+ Hetero. + Variational Dropout (I) &\cellcolor{blue!15} $6.291\pm0.012$&$13.906\pm0.048$ & \cellcolor{red!15}$\mathbf{8.944\pm0.044}$ &$16.761\pm0.047$\\ 
%		+ Hetero. + Variational Dropout (II) &\cellcolor{red!15} $\mathbf{6.287\pm0.029}$&$13.927\pm0.093$&\cellcolor{blue!15} $8.955\pm0.029$&$16.844\pm0.109$\\
		\hline
	\end{tabular}%	
	\vspace{-1mm}
\caption{Super-resolution results on diffusion tensor images (DTIs) of HCP and Lifespan datasets for different upsampling methods. For each method, an ensemble of $10$ models are trained on different training sets generated by randomly extracting a set of patch pairs from the common $16$ HCP subjects. For each model, the average RMSE ($\times 10^{-4} \text{mm}^2/\text{s} $) over subjects in respective datasets is first computed and the mean/std of such average RMSE over the ensemble are then reported. Best results in bold red, and the second best in blue. }
\label{tab:compare_1}	
\end{table}

\begin{table}
	\center
	\footnotesize
	\begin{tabular}{@{}lclclclcl}
		\hline
		Models & HCP (interior) & HCP (exterior)& Life (interior) & Life (exterior)   \\ 
		\hline	
		CSpline interpolation& $5.234 \pm$ n/a & $30.362\pm$ n/a & $7.135\pm$ n/a & $29.232\pm$ n/a  \\
		$\beta$-Spline interpolation &  $4.852 \pm$ n/a  &  $63.446\pm$ n/a & $6.523\pm$ n/a &  $56.937\pm$ n/a \\
		IQT-RF \cite{alexander2017image}&   $4.538 \pm 0.113$& $25.541\pm0.131$  & $5.882\pm0.121$  & $26.137\pm0.279$   \\ 
		BIQT-RF \cite{tanno2016bayesian} & $4.838 \pm 0.129$  &$25.523\pm0.175$  & $5.949\pm0.131$ &$27.509\pm0.233$ \\
				
		\hline
		3D-ESPCN(baseline)  & $2.285 \pm0.126$   & $9.316\pm0.127$ & $4.195 \pm0.163$ & $11.922 \pm0.192$ \\
		+ Binary Dropout ($p=0.1$) &    $2.283 \pm0.154$	 & $9.272\pm0.132$ & $4.120 \pm0.178$ &	$11.652 \pm0.204$	\\
		+ Gaussian Dropout ($p=0.1$)  &    $2.370 \pm0.155$	& 	$9.335\pm0.144$	& 	$4.327\pm0.157$	&	$11.907 \pm0.211$	\\
		+ Variational Dropout (I)& $2.155\pm0.122$ &   $9.205\pm0.193$ 	& 	$3.997 \pm0.153$  & $11.547 \pm0.177$ 	\\
		+ Variational Dropout (II)& $2.172\pm0.128$ &  $9.112\pm0.173$  &  $3.972 \pm0.132$ &  $11.511 \pm0.172$ & \\
		+ Hetero. &$1.998\pm0.132 $ & $11.294\pm 0.216$  & $3.872 \pm0.140$   & $12.084 \pm0.129$\\
		+ Hetero + Variational Dropout (I) &\cellcolor{red!15} $\textbf{1.951} \pm \textbf{0.122}$ & \cellcolor{blue!15}$9.102\pm0.181$ & \cellcolor{red!15} $\textbf{3.572}\pm\textbf{0.171}$  &\cellcolor{red!15}$\textbf{11.037} \pm\textbf{0.192}$  \\ 
		+ Hetero + Variational Dropout (II)&\cellcolor{blue!15} $1.969 \pm 0.119$ &\cellcolor{red!15}$\textbf{9.052}\pm\textbf{0.162}$& \cellcolor{blue!15} $3.606 \pm0.141$ & \cellcolor{blue!15}$11.311 \pm0.195$  \\
		\hline
		3D-ESPCN(without outlier removal)  & $3.425 \pm0.163$   & $13.284\pm0.239$ & $6.032 \pm0.229$ & $15.5 13 \pm0.273$ \\
		+ Hetero. &$2.264\pm0.153 $ & $11.306\pm 0.172$  & $3.919 \pm0.140$   & $12.821 \pm0.150$\\
		+ Hetero + Variational Dropout (I) & $2.138 \pm0.159$ & $10.022\pm0.187$ &  $3.681\pm0.193$  &$12.133 \pm 0.205 $  \\ 
		+ Hetero + Variational Dropout (II)& $2.133 \pm 0.188$ & $ 9.988 \pm 0.209$&  $3.690 \pm0.184$ & $12.052 \pm0.212$  \\
		\hline
	
	\end{tabular}%	
\vspace{-1mm}
\caption{Super-resolution results on MAP-MRIs of HCP and Lifespan datasets for different upsampling methods. For each method, an ensemble of $5$ models are trained on different training sets generated by randomly extracting a set of patch pairs from the common $16$ HCP subjects. For each model, the average RMSE over subjects in respective datasets is first computed and the mean/std of such average RMSEs over the ensemble are then reported. Best results in bold red, and the second best in blue. In addition, the performance of 3D-ESPCN and its probabilistic variants trained on data without outlier removal are also included.}
\label{tab:compare_2}
\end{table}

% Table for 8 training subjects 
%\begin{figure}[t]
%	%\vspace{- 15pt}
%	\centering
%	\includegraphics[width=\linewidth]{figures/table_1_3.pdf}\label{fig:performance_compare}
%	\small
%	\caption{(a) RMSE on HCP and Lifespan dataset for different upsampling methods. For each method, an ensemble of $10$ models are trained on different training sets, and the mean/std of the average errors over $8$ test subjects are computed over the ensemble. Best results in bold red, and the second best in blue. (b) Interior (yellow) and exterior region (red).} 
%	%\vspace{- 15pt}
%\end{figure}

Table~\ref{tab:compare_1} shows that, on both HCP and Lifespan data, modelling both intrinsic and parameter uncertainty (i.e. Hetero. + Variational Dropout (I), (II)) achieves the best reconstruction accuracy in DTI super-resolution. We observe that modelling intrinsic uncertainty with the heteroscedastic network on its own further reduces the average RMSE of the baseline 3D-ESPCN on the interior region with high statistical significance ($p<10^{-3}$). However, poorer performance is observed on the exterior than the baseline. On the other hand, using $200$ MC weight samples, we see modelling parameter uncertainty with variational dropout (see Variational Dropout.(I)-CNN) performs best on both datasets on the exterior region. Combination of heteroscedastic model and variational dropout (i.e. Hetero. + Variational Dropout (I) or (II)) leads to the top 2 performance on both datasets on the interior region and reduces errors on the exterior to the level comparable or better than the baseline. 

%Table ~\ref{tab:compare_1} shows that, on both HCP and Lifespan data, the heteroscedastic network further reduces the average RMSE of the baseline 3D-ESPCN on DTI super-resolution on the interior region with high statistical significance ($p<10^{-3}$). However, poorer performance is observed on the exterior than the baseline. Using $200$ MC weight samples, we see Var.(I)-CNN performs best on both datasets on the exterior region. Combination of heteroscedastic model and variational dropout (i.e. Hetero+Var.(I) or (II)) leads to the top 2 performance on both datasets on the interior region and reduces errors on the exterior to the level comparable or better than the baseline. 

Similarly, Table~\ref{tab:compare_2} shows that the best performance in MAP-MRI super-resolution comes from the combined models (i.e. Hetero.+Variational Dropout.(I) and (II)). We observe that as with the DTI case,  modelling intrinsic uncertainty through the heteroscedastic network improves the reconstruction accuracy on the interior region, whilst the errors on the exterior are increased with respect to the baseline 3D-ESPCN. Moreover, the improvement is pronounced when the outliers due to model fitting errors are not removed in the training data. In this case, we see that the reconstruction accuracy of 3D-ESPCN dramatrically decreases, whilst in contrast it is only marginally compromised when equipped with the heteroscedastic noise model, displaying robustness to outliers. Lastly, we note that the top-2 accuracy are consistently achieved by the joint modelling of intrinsic and parameter uncertainty (i.e. Hetero.+Variational Dropout.(I) and (II)) on both the interior and exterior regions on both HCP and Lifespan datasets.  

%Table~\ref{tab:compare_2} compares the similar results for MAP-MRI super-resolution. We observe that as with the DTI case,  the heteroscedastic network improves the reconstruction accuracy on the interior region, whilst the errors on the exterior are increased with respect to the baseline 3D-ESPCN. However, when the outliers due to model fitting errors are not removed in the training data, the reconstruction accuracy of 3D-ESPCN dramatrically decreases, whereas it is only marginally compromised when equipped with the heteroscedastic noise model. Notably, the top-2 accuracy are achieved by the combined models (i.e. Hetero+Var.(I) and Hetero+Var.(II)) on both the interior and exterior regions on both HCP and Lifespan datasets. 

The performance difference of heteroscedastic network between the interior and the exterior region roots from the loss function. The Mahalanobis term $\mathcal{M}_{\theta}(\mathcal{D})$ in eq.\eqref{eq:loss_hetero} imposes a larger penalty on the regions with smaller intrinsic uncertainty. The network therefore allocates less of its resources towards the regions with higher uncertainty (e.g. boundary regions) where the statistical mapping from the low-resolution to high-resolution space is more ambiguous, and biases the model to fit the regions with lower uncertainty. However, we note that the performance of the heteroscedastic network is still considerably better than the standard interpolation and RF-based methods. By augmenting the model with variational dropout, the exterior error of the heteroscedastic model is dramatically reduced, indicating its regularisation effect against overfitting to low-uncertainty areas. We also observe concomitant performance improvement on the interior regions on both datasets, which additionally shows the benefits of such regularisation even in low-uncertainty areas. 

Both Table~\ref{tab:compare_1} and Table~\ref{tab:compare_2} show that the use of variational dropout attains lower errors than the models with fixed dropout probabilities $p$, namely, Binary and Gaussian dropout \cite{srivastava2014dropout}. Different instances of both dropout models are trained for a range of $p$ by linearly increasing on the interval $[0.05,0.3]$ with increment $0.05$, and the test errors for the configurations with smallest RMSE on the validation set are reported in Table~\ref{tab:compare_1} and Table~\ref{tab:compare_2}. As with variational dropout models, $200$ MC samples are used for inference. In all cases, two variants of variational dropout (I) and (II) outperform the networks with the best binary or Gaussian dropout models, showing the benefits of learning dropout probabilities $p$ rather than fixing them in advance. 

\subsection{Tractography with MAP-MRI}
Reconstruction accuracy does not necessarily reflect real world utility. We thus further assessed the benefits of super-resolution with a tractography experiment on the Prisma dataset, which contains two DWIs of the same subject at two different image resolutions---1.35 mm and 2.5 mm isotropic voxels, as detailed in Sec.~\ref{sec:dataset}. An ensemble of $10$ best performing CNN (3D-ESPCN+Hetero.+Variational Dropout(I)) is used to super-resolve the MAP-MRI coefficients \cite{ozarslan2013mean} derived from the low-resolution DWIs, and the ensemble predictions aggregated into the final output by taking the average estimate weighted by the inverse of the estimated intrinsic uncertainty. Lastly, the high-resolution multi-shell DWIs are obtained from this super-resolved MAP volume. Specifically, the Spherical Mean Technique (SMT) is used to fit a microscopic tensor model to the predicted dataset \cite{kaden2016quantitative}. The voxel-by-voxel estimated model parameters inform the spatially varying fibre response function that is used to recover the fibre orientation distribution through spherical deconvolution. Afterwards, we perform probabilistic tractography \cite{tournier2010improved} with the fibre pathways randomly seeded in the brain. In a similiar fashion, we also generate high-resolution datasets by using IQT-RF and linear interpolation.

%(Ryu): Reconstruction accuracy does not reflect real world utility, thus we further assessed the benefits of super-resolution with a downstream tractography experiment on the Prisma dataset, which contains two DWIs of the same subject from a Siemens Prisma 3T scanner, at two different image resolutions (1.35 mm and 2.5 mm isotropic voxels). The full details of the dataset are available in Sec.~\ref{sec:dataset}. An ensemble of $8$ 3D-ESPCN+Hetero.+Variational Dropout(I) is used to super-resolve the MAP-MRI coefficients \cite{ozarslan2013mean} derived from the low-resolution DWIs (2.5 mm), and the ensemble predictions aggregated into the final output by taking the average estimate weighted by the inverse of the estimated intrisic uncertainty. Lastly, the high-res DWIs (1.25 mm) are obtained from this super-resolved MAP volume. In a similiar fashion, we also generate high-res datasets by using IQT-RF and linear interpolation.

\begin{figure}[ht]
	\vspace{3mm}
	\centering
	\includegraphics[width=\linewidth]{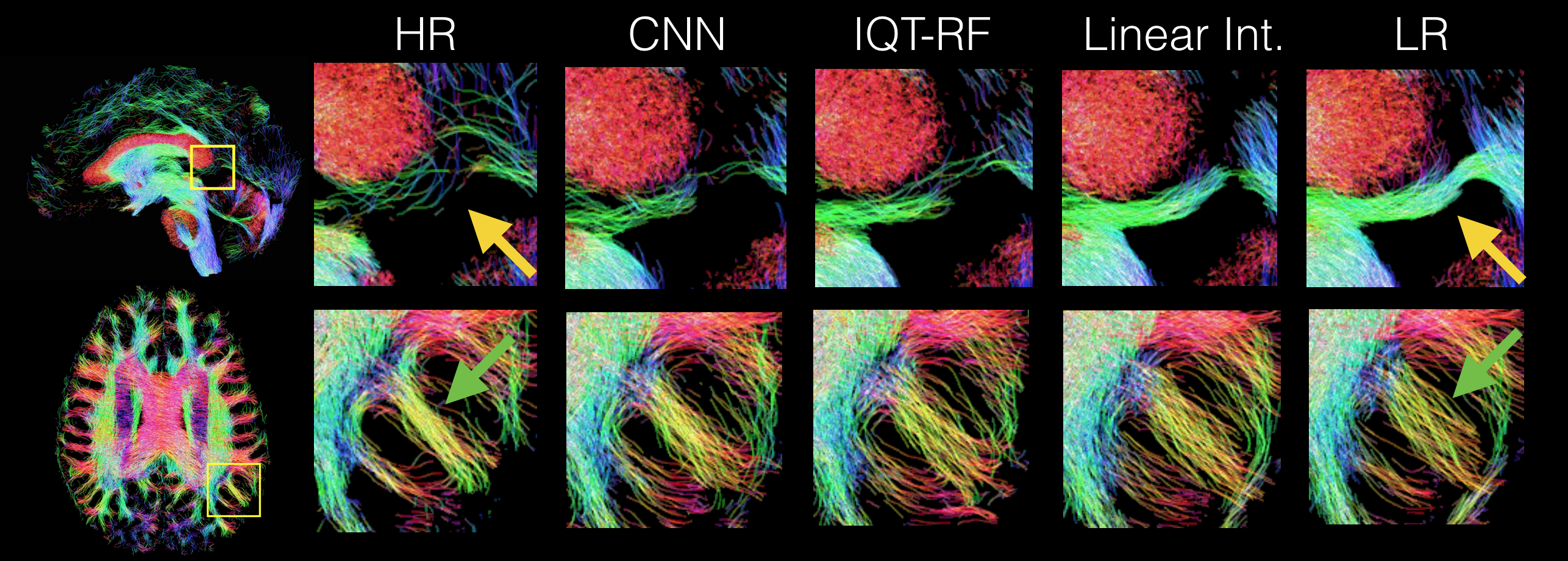}\label{fig:uncertainty_map}
	\small
	\caption{Streamline maps of the probabilistic tractography \cite{tournier2010improved} applied on Prisma dataset for different upsampling methods and visualised with MRtrix3 \cite{tournier2012mrtrix}. From left to right: (i) High-res acquisition, (ii) CNN (3D-ESPCN+Hetero.+Variational Dropout(I)) prediction; (iii) Random Forest IQT \cite{alexander2017image}; (iv) Linear interpolation; (v) Low-res acquisition. The yellow arrows in the top row indicate the location of a false positive tract detected in the low-resolution acquisition, whilst the green arrows in the bottom row show an example white matter tract, which is more sharply reconstructed at high resolution. } 
	\label{fig:tract}
	%	\vspace{-1em}
\end{figure}

\clearpage
Fig.~\ref{fig:tract} shows that IQT via our best performing CNN makes a tangible difference in downstream tractography. In the top row, tractography on the low-resolution data produces a false-positive tract under the corpus callosum (yellow arrow), which tractography at high resolution avoids. Reconstructured high-resolution images from IQT-RF and CNN predictions avoid the false positive better than linear interpolation. Note that we do not expect to reproduce the high-resolution tractography map exactly, as the high-resolution and low-resolution images are not aligned exactly and the high-resolution and prediction have different resolutions (1.35 mm vs. 1.25 mm). The bottom row shows sharper recovery of small gyral white matter pathways (green arrow) at high-resolution than low-resolution resulting from reduced partial volume effect. CNN reconstruction produces a sharper pathway than RF-IQT and linear interpolation, more closely reflecting the high-resolution tractography.
%In the top row, tractography on the low-resolution data indicates presence of a false-positive tract under the corpus callosum (yellow arrows) while not picked up in the high-resolution data. Both IQT-RF and CNN predictions appear less sensitive to false indication tracts in the input compared to linear interpolation. Also, the bundle structures close to the outer boundary (green arrow) gracefully becomes shaper moving right to left. This is consistent with the CNN-based improvements on the exterior region for DTI-SR.

\subsection{Uncertainty Quantification}
%\textcolor{red}{Need to sharpen the preamble to better motivate the experiments we perform if necessary. Or we could remove this and motivate better at the start of the respective subsections \ref{sec:uncertainty_HCP} and \ref{sec:unseen_abnormality}.}
 In this section, we investigate the value of uncertainty modelling in enhancing the safety of super-resolution system beyond reduced reconstruction errors. Firstly, in Sec.~\ref{sec:uncertainty_HCP}, we study the utility of predictive uncertainty map as a proxy measure of reconstruction accuracy on healthy test subjects from both HCP and Lifespan datasets. Secondly, in Sec.~\ref{sec:unseen_abnormality}, we look into the behaviour of uncertainty maps in the presence of abnormal features that are not present in the training data. 

% In particular, we aim to demonstrate that the proposed uncertainty quantification method not only provides a subject-specific and spatially varying metric of output reliability, but also adds a high-level transparency to the predictive performance by decoupling the sources of predictive uncertainty. 

%In this section, we aim to display the value of uncertainty modelling beyond performance imrovement. In particular, we demonstrate quantification of uncertainty not only provides a safer means to deploy the super-resolution system, but also adds a high-level transparency to the predictive performance through decoupling the sources of uncertainty. 

\subsubsection{Healthy Test Subjects} \label{sec:uncertainty_HCP}
We employ the most performant CNN model (3D-ESPCN + Hetero. + Variational Dropout(I)) to generate the high-resolution predictions of \textit{mean diffusivity} (MD) and \textit{fractional anisotropy} (FA), and their associated predictive uncertainty maps.  Here we draw $200$ samples of high-resolution DTI predictions for each subject from the predictive distribution $q_\phi^*(\mathbf{y}|\mathbf{x})$, and then the FA and MD maps of each prediction are computed. The sample mean and standard deviation are then calculated from these samples to generate the final estimates of high-resolution MD/FA maps and their corresponding predictive uncertainty. 

Fig.~\ref{fig:uncertainty_map} displays high correspondence between the error (RMSE) maps and the predictive uncertainty on both FA and MD of a HCP test subject. This demonstrates the potential utility of uncertainty map as a surrogate measure of prediction accuracy. In particular, the MD uncertainty map captures subtle variations within the white matter and the cerebrospinal fluid (CSF) at the centre. Also, in accordance with the low reconstruction accuracy, high predictive uncertainty is observed in the CSF in MD. This is expected since the CSF is essentially free water with low signal-to-noise-ratio (SNR) and is also affected by biological noise such as cardiac pulsations. The reconstruction errors are high in FA prediction on the bottom-right quarter of the brain boundary, close to the skull, which is also reflected in the uncertainty map. 

\begin{figure}[t]
	%\vspace{-2em}
	
	\centering
	\includegraphics[width=\linewidth]{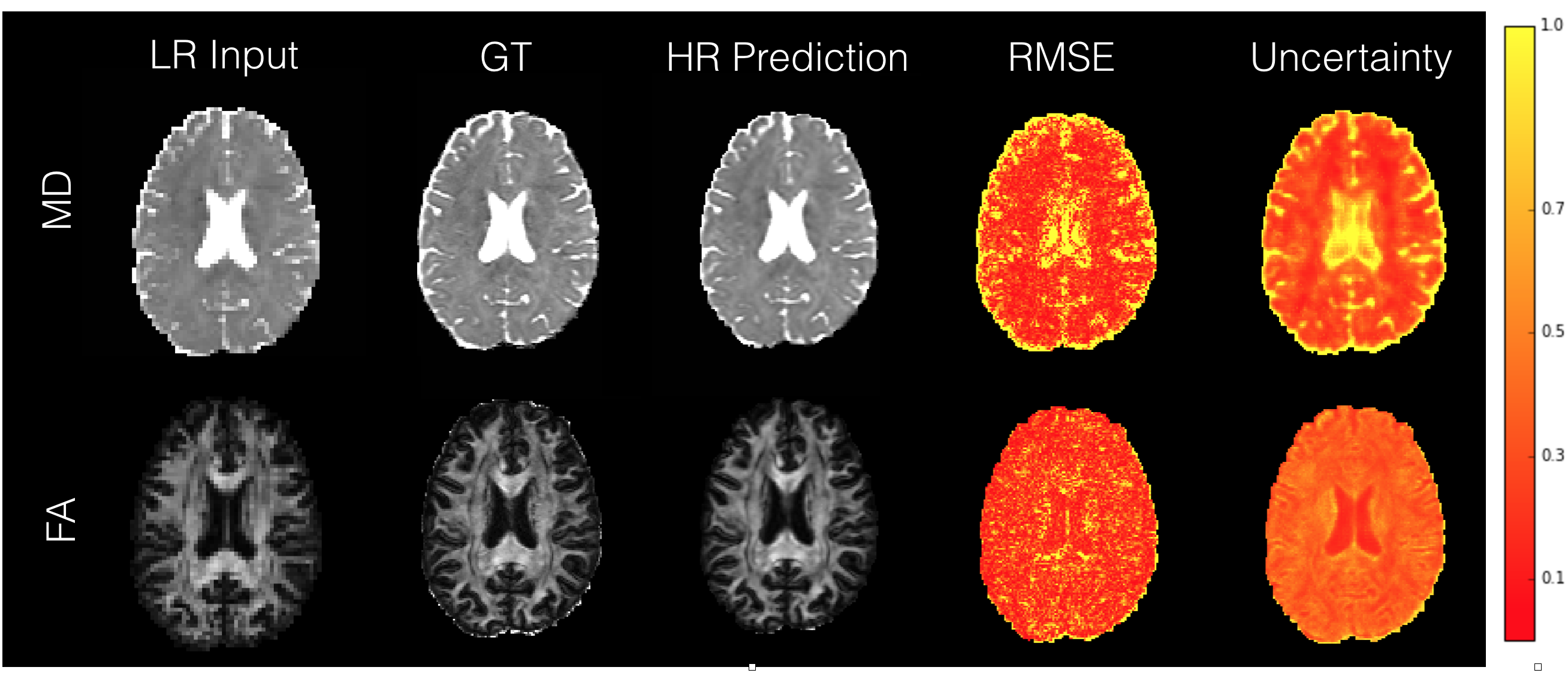}
	\small
	\caption{Comparison between voxel-wise RMSE and predictive uncertainty maps for FA and MD computed on a HCP test subject (min-max normalised for MD and FA separately). Low-res input, ground truth and the mean of high-resolution predictions are also shown.} 
	\label{fig:uncertainty_map}
	%	\vspace{-1em}
\end{figure}

\begin{figure}[t] 
	\centering
	\includegraphics[width=0.8\linewidth]{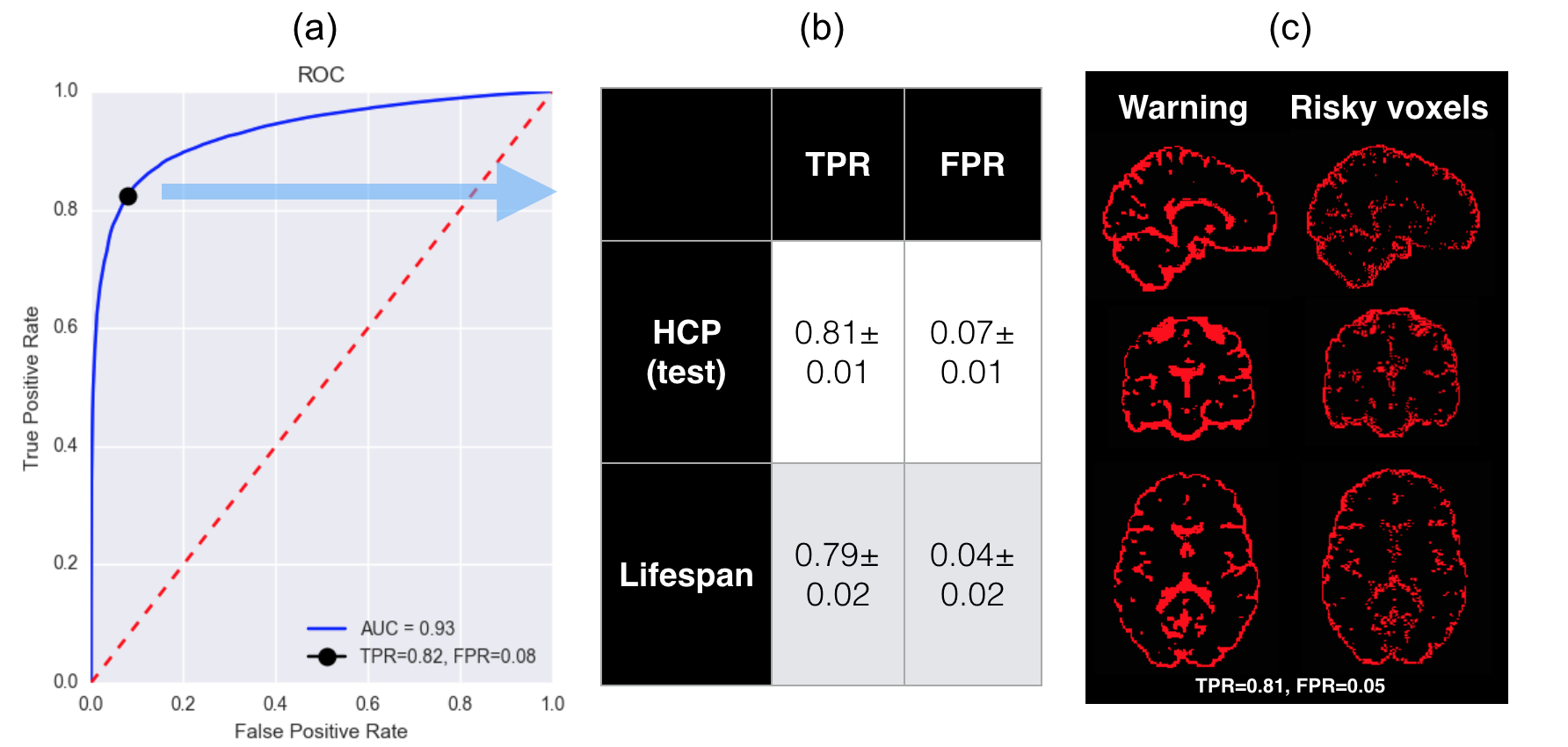}
	\small
	\caption{Discrimination of ``safe'' voxels in the predicted high-resolution MD map by thresholding on predictive uncertainty. Here a single 3D-ESPCN + Hetro. + Variational Dropout (I) model is used to quantify the predictive uncertainty over each image volume. (a) the ROC curve plots the true positive rate (TPR) against false positive rate (FPR) computed  for a range of threshold values on the foreground voxels in the training subjects. Best threshold (black dot) was selected such that F1 score is maximised and is employed to separate ``safe'' voxels from ``risky'' ones; (b) the average TPR and FPR over the 16 test HCP subjects and the 16 Lifespan subjects are shown; (c) an example visualisation of the ``ground truth'' safe (black) and risky (red) voxels on a Lifespan subject along with the corresponding classification results denoted as ``warning''. }

	\label{fig:roc}
\end{figure}

Fig.~\ref{fig:roc} tests the utility of predictive uncertainty map in discriminating potential predictive failures in the predicted high-resolution MD map. We define ground truth ``safe'' voxels as the ones with reconstruction error (RMSE) smaller than a fixed value, and the task is to separate them from the remaining ground-truth ``risky'' voxels  by thresholding on their predictive uncertainty values. The threhold for defining safe voxels is set to $1.5\times10^{-4} \text{ s/mm}^2$, such that the risky voxels mostly concentrate on the outer-boundary and the CSF regions (which account for $17.5\%$ of all voxels under consideration). Here the positive class is defined as ``safe'' while  the negative class is defined as ``risky''. Fig.~\ref{fig:roc} (a) shows the corresponding receiver operating characteristic (ROC) curve of such binary classification task, which plots the true-positive-rate (TPR) against the false-positive-rate (FSR) computed based on all the voxels in the 16 HCP training subjects. In this case, TPR decribes the percentage of correctly detected safe voxels out of all the safe ones, while FPR is defined as the percentage of risky voxels that are wrongly classified as safe out of all the risky voxels. We then select the best threshold by maximising the F1 score, and use this to classify the voxels in each predicted high-resolution MD into ``safe'' and ``risky'' ones for all subjects in the test HCP dataset and the Lifespan dataset. Fig.\ref{fig:roc} (b) shows the inter-subject average of the TPR and FPR on both datasets. While on average TPR slightly worsens compared to the results on the training subjects, FPR improves in both cases---notably, this uncertainty-based classification is able to correctly identify 96\% of risky predictions on unseen subjects from out-of-training-distribution dataset, namely Lifespan, which differs in demographics and underlying acquisition. Fig.\ref{fig:roc} (c) visualises the classification results to the pre-defined ``ground truth" on one of the Lifespan subjects, which illustrates that the generated ``warning'' aggressively flags potentially risky voxels at the cost of thresholding out the safe ones. 

%We define the ‘safe’ voxels as the ones with reconstruction error (RMSE) smaller than a fixed value, and want to separate them from the rest by thresholding on their predictive uncertainty values. The threhold for defining ‘ground truth’ safe voxels is set to $1.5\times10^{-4} \text{ s/mm}^2$, such that the non-safe voxels mostly concentrate on the outer-boundary and the CSF regions (which accout for $17.5\%$ of all voxels under consideration). The receiver operating characteristic (ROC) curve of such binary classification task is then plotted based on the all the voxels in 16 of the HCP training subjects and the best threshold is chosen as the one with the maximal F1 score with the false positive rate below $10\%$ (see Fig.\ref{fig:roc} (a)). Then, this optimal threshold is used to classify the voxels in each predicted high-resolution MD into ‘safe’ (black) and ‘risky’ ones (red) for all subjects in the test HCP dataset and the Lifespan dataset. Fig.\ref{fig:roc} (b) shows the inter-subject average of the true-positive-rate (TPR) and false-positive-rate (FPR) on both datasets. While on average TPR slightly worsens, FPR improves in both cases---notably, this uncertainty-based classification is able to correctly identify 96\% of risky predictions on unseen subjects from out-of-training-distribution dataset, namely Lifespan. Fig.\ref{fig:roc} (c) visualises the classification results to the pre-defined ``ground truth" on one of the Lifespan subjects.

%In particular, we applied the best-performing SR model (Hetero+Variational (II) ESPCN) trained on a healthy HCP cohort to super-resolve the DTI of a brain tumour patient. The raw data (DWI) is processed as before, and the $b = 700 \text{ s/mm}^2$ measurement with voxel size  $2^3 \text{ mm}^3$ is used as the input. The observation and model uncertainty on the first DTI component are separately illustrated along with the input and the predicted high-resolution image. Although the ground truth is unavailable, the estimated image seems to sharpen the original image without introducing noticeable spurious features. The observation uncertainty map indicates higher uncertainty to the region of tumour than the surrounding healthy white matter as desired. Understanding the behaviours of these uncertainty measures in an `unfamiliar' test environment and their relations to predictive performance remains an important future work for designing a more generalisable method.

\subsubsection{Unseen Abnormalities and Uncertainty Decomposition} \label{sec:unseen_abnormality}
We separately visualise the propagated intrinsic and parameter uncertainty over the predicted high-resolution MD map on images of subjects with a variety of different unseen abnormal structures, such as benign cysts, tumours (Glioma) and focal lesions caused by multiple sclerosis (MS). We emphasise here that the all these images have been acquired with different protocols. Specifically, benign cysts in the HCP datasets represent abnormalities in images acquired with the same protocol as the training data, while tumours and MS lesions are examples of pathologies present in out-of-distribution imaging protocols. In all cases, we use the SR network, Hetero.+Variational Dropout (I), trained on healthy subjects from HCP dataset. For each of $200$ different sets of parameters $\{\theta_{t}\}_{t=1}^{200}$ sampled from the posterior distribution $q(\theta|\mathcal{D})$,  we draw $10$ samples of high-resolution DTIs from the likelihood, $\{\mathbf{y}^{t}_{j}\}_{j=1}^{10} \sim p(\mathbf{y}|\theta_{t},\mathbf{x},\mathcal{D})$, compute the corresponding MD, and approximate the two constituents of predictive uncertainty with the MC estimators given in eq.\eqref{eq:mc_model_uncertainty} and \eqref{eq:mc_intrinsic_uncertainty}. 

\begin{figure}[ht]
	\centering
	\subfigure[HCP subject with a benign cyst]{\includegraphics[width=0.7\linewidth]{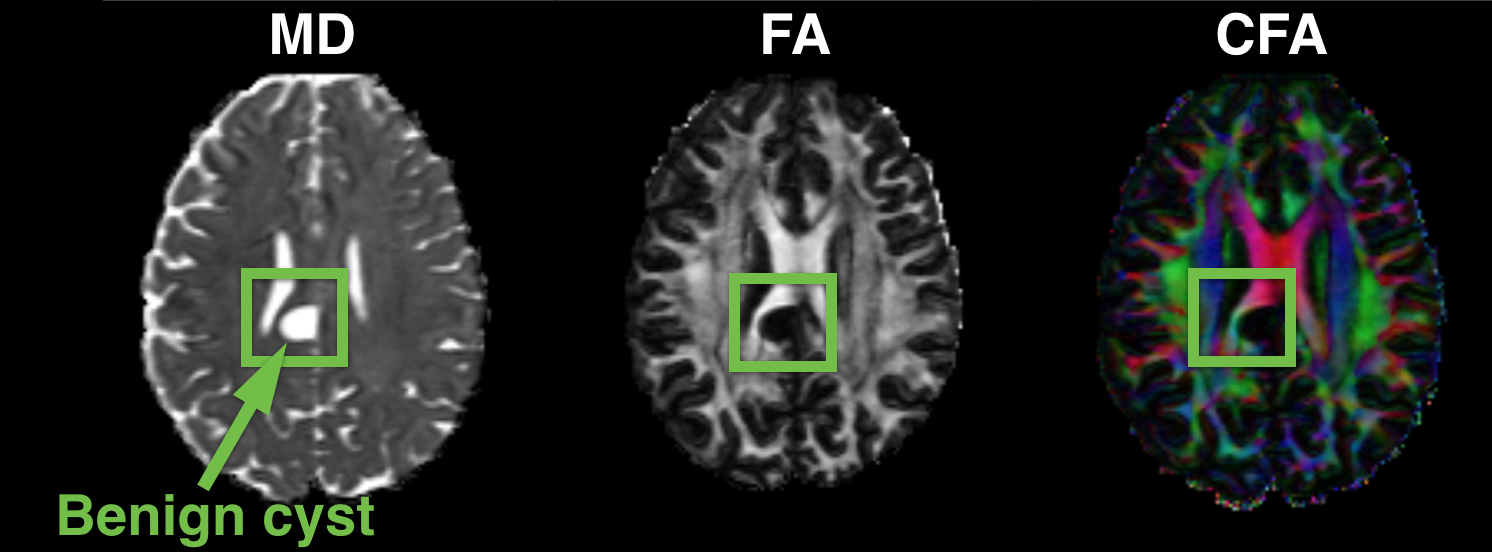}}
	\subfigure[Errors vs. uncertainty components ]{\includegraphics[width=0.7\linewidth]{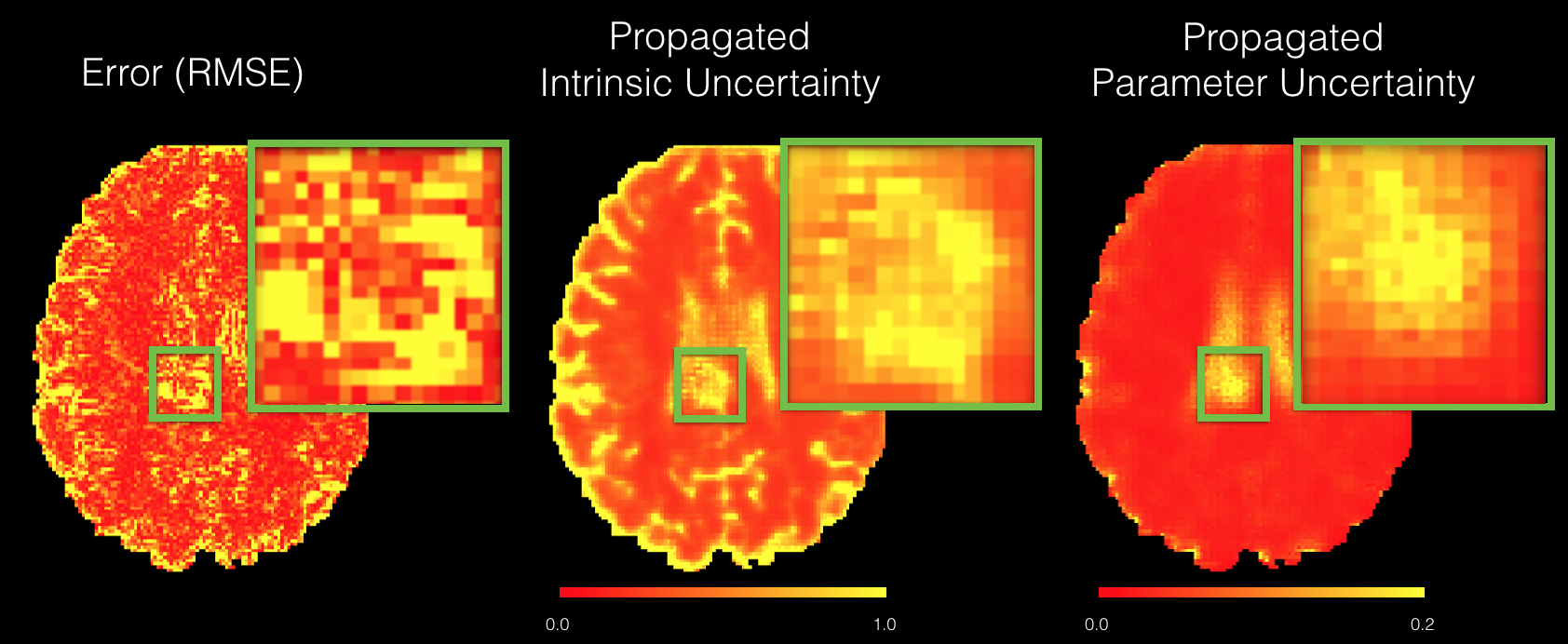}}
	\caption{Visualisation of (a) MD, FA and colour FA maps computed from the DTI of a HCP subject with a small posterior midline arachnoid cyst in the central part of the brain. (b) the corresponding reconstruction accuracy (RMSE) in MD and the corresponding components of predicted uncertainty. }
	\label{fig:healthy_abnormal}
\end{figure}

Fig.~\ref{fig:healthy_abnormal} shows the reconstruction accuracy along with the components of predictive uncertainty over the high-resolution MD map of a HCP test subject, which contains a benign abnormality (a small posterior midline arachnoid cyst). The error (RMSE) and propagated intrinsic uncertainty are plotted on the same scale whereas the propagated model uncertainty is plotted on 1/5 of the scale for clear visualisation. In this case, the predictive uncertainty is dominated by the intrinsic component. In particular, low propagated intrinsic uncertainty is observed in the interior of the cyst relative to its boundary in accordance with the high accuracy in the region. This is expected as the interior structure of a cyst is highly homogeneous with low variance in signals and the super-resolution task should therefore be relatively straightforward. On the other hand, the component of parameter uncertainty is high on the interior structure which also makes sense as such homogeneous features are underrepresented in the  training data of healthy subjects. This example illustrates how decoupling the effects of intrinsic and parameter uncertainty potentially allows one to make sense of the predictive performance. 

\begin{figure}[ht] 
	\centering
	\includegraphics[width=\linewidth]{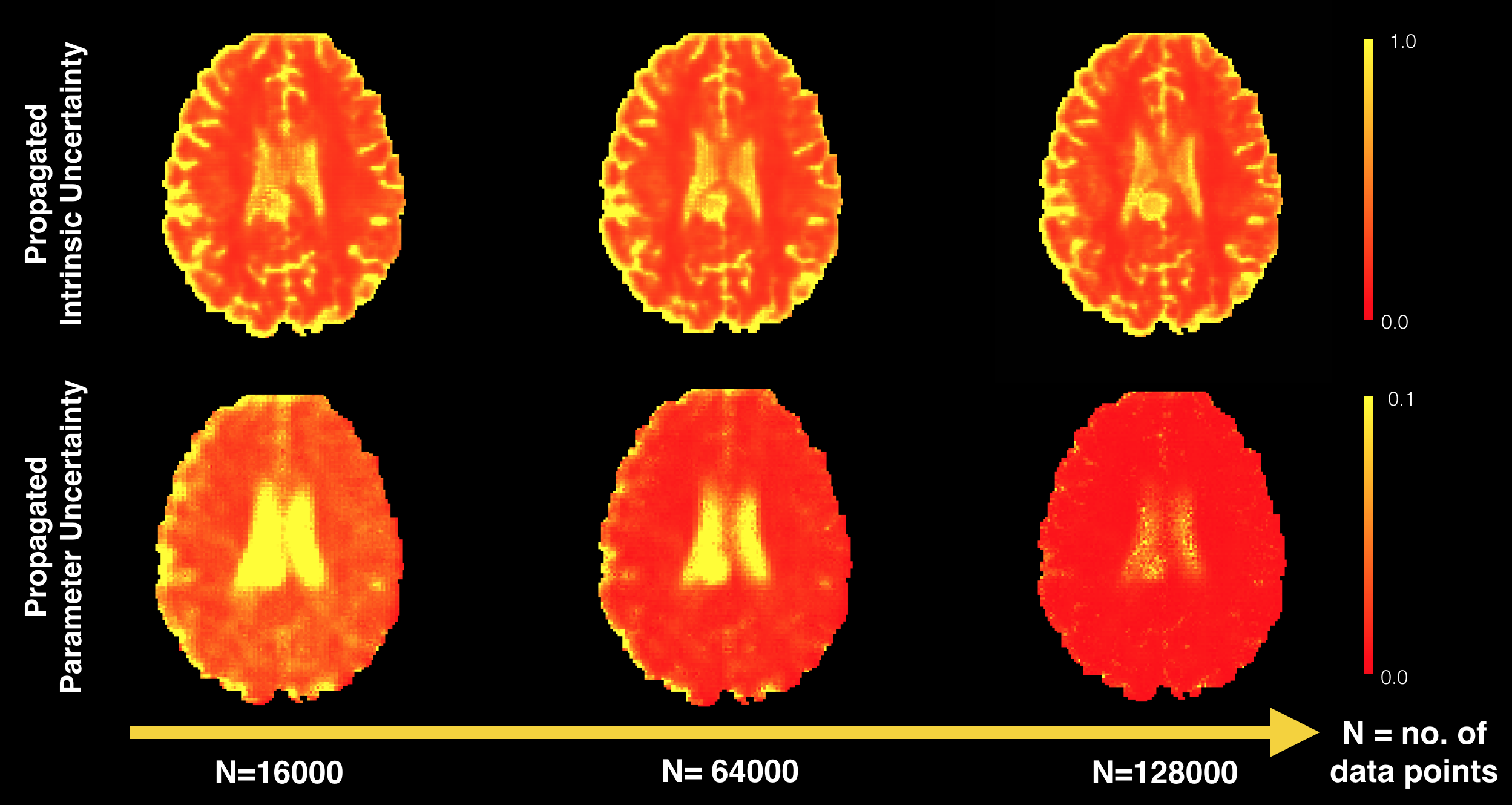}
	\small
	\caption{Training set size vs propagated intrinsic/parameter uncertainty on the MD map of an unseen HCP subject with a benign cyst. The uncertainty maps are normalised across all the figures. } 
	\label{fig:uncertainty_decomp_1}
\end{figure}

Fig.\ref{fig:uncertainty_decomp_1} visualises the uncertainty components generated by the same CNN model trained on datasets of varying size. We see that the propagated parameter uncertainty diminishes as the training set size increases, while the propagated intrinsic uncertainty stays more or less constant. This result is indeed what is expected as described in Fig.~\ref{fig:uncertainty_types}; the specification of network weights becomes more confident i.e. the variance of the posterior distribution decreases as the amount of training data increases, while the effect of intrinsic uncertainty is irreducible with the amount of data.  On the other hand, when the standard binary or Gaussian dropout was employed instead of variational dropout, we observed that the effect of parameter uncertainty stayed more or less constant with the size of training data. This may be a consequence of the posterior variance, largely determined by the prespecified drop-out rates, which in turn results in more static variance of predictive distribution. 

%\begin{figure}[ht] 
%	\centering
%	\includegraphics[width=0.65\linewidth]{figures/fig_12.png}
%	\small
%	\caption{MD, FA and colour FA maps computed from the DTI of a HCP subject ($105620$) with a small posterior midline  arachnoid cyst in the central part of the brain.} 
%	\label{fig:healthy_abnormal}
%\end{figure}

%
%\begin{figure}[ht] 
%	\centering
%	\includegraphics[width=\linewidth]{figures/fig_9_2.png}
%	\small
%	\caption{Comparison between reconstruction accuracy and components of predicted uncertainty on a HCP subject with a benign abnormality. } 
%	\label{fig:uncertainty_decomp_2}
%\end{figure}
\begin{figure}[ht]
	\centering
	\subfigure[Brain Tumour]{\includegraphics[width=\linewidth]{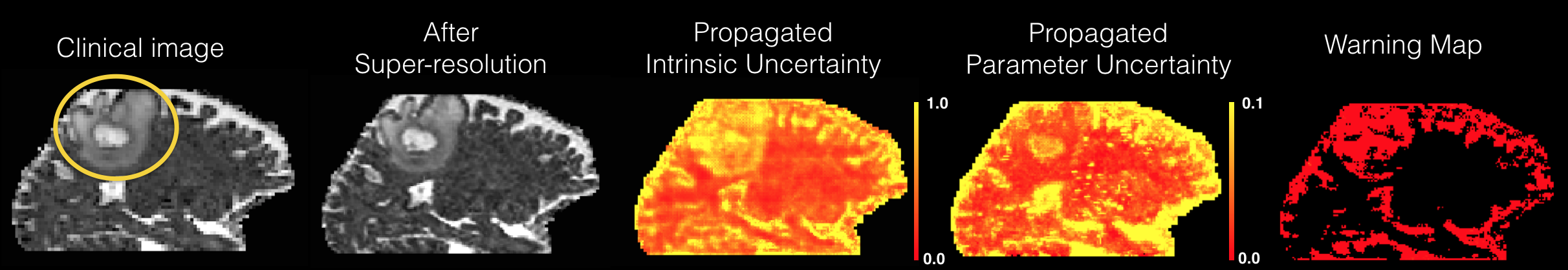}}
	\subfigure[Multiple Sclerosis ]{\includegraphics[width=\linewidth]{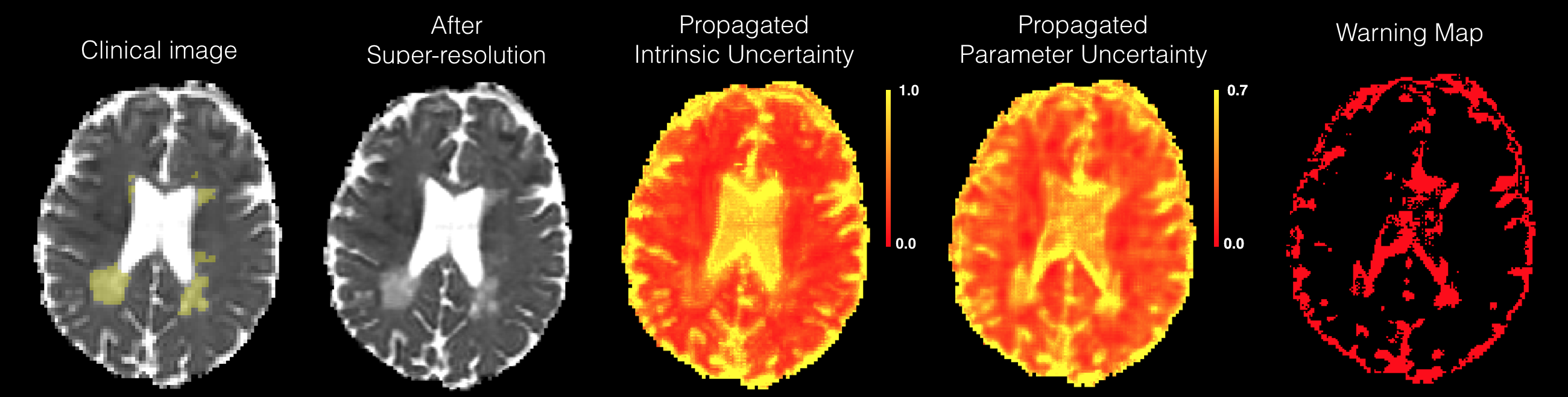}}
	\caption{Visualisation of propagated uncertainty components on clinical images with pathology that was not present in the training data. The super-resolution is performed on the clinical images due to low-resolution, and thus the ground truths are not available in both cases. (a) shows the results on the data of a Glioma patient, and the yellow circle indicates the region of tumour. (b) shows the same set of results on a MS patient with labels of focal lesions obtained from a neurologist indicated in yellow. Each row shows from left to right: (i) MD map computed from the original DTI; (ii) MD map computed from the output of super-resolution; (iii), (iv) maps of the estimated propagated intrinsic and parameter uncertainty; (v) ``warning map'' obtained from the same threshold value used in Sec.~\ref{sec:uncertainty_HCP}, which flag large parts of the pathological features in both cases. }
	\label{fig:components_pathology}
\end{figure}

We further validate our method on clinical images with previously unseen pathologies. We note that the pathology data contain images acquired with standard clinical protocols with voxel size slightly smaller than that of the training low-resolution images and lower signal-to-noise ratio. 

%\begin{figure}[t]
%	%\vspace{-2em}
%	\centering
%	\includegraphics[width= 0.25\linewidth]{figures/fig_6_2.png}
%	\small
%	\caption{DTI super-resolution on the data of a brain tumour patient. From top to bottom: (i) MD computed from the original DTI; (ii) the output of super-resolution; (iii) predictive uncertainty. In total, $200$ samples are drawn from the network's predictve distribution to compute (ii) and (iii). } 
%	\label{fig:tumour}
%	%	\vspace{-1em}
%\end{figure}

Fig.~\ref{fig:components_pathology} shows that pathological areas not represented in the training set are flagged as highly uncertain. Although the ground truth is not available in this case, the uncertainty can be quantified instead to flag potential low accuracy areas. Fig.~\ref{fig:components_pathology} (a) shows that the propagated parameter uncertainty highlights the tumour core, and speckly artefacts in the input image, which are not represented in the training data. On the other hand, the intrinsic uncertainty component is high on the whole region of pathology covering both the tumour core and its surrounding edema. Fig.~\ref{fig:components_pathology} (b) shows that high parameter uncertainty is assigned to a large part of focal lesions in MS, while the intrinsic uncertainty is mostly prevalent around the boundaries between anatomical structures and CSF. We also observe that the super-resolution sharpens the original image without introducing noticeable artifacts; in particular, for the brain tumour image, some of the partial volume effects are cleared. 
%\newline 
%\\\textcolor{red}{TODO: it would be nice to demonstrate on different types of pathology e.g. a variety of healthy abnormal features in HCP subjects, MS lesions. }
%\newline
%\\\textcolor{red}{Update 15/Dec/2017: experiments show that the uncertainty maps do not highlight the lesions as clearly as for brain tumours. However, this may be indeed the correct indication i.e. the performance on lesions are not too bad, which is unfortunately not possible to test with the current dataset. Experiments on abnormal HCP subjects suggests that low-uncertainty occasionally assigned to abnomalities seem to correlate with the errors. }

\section{Discussion and Conclusion}
% NOTES and CONTENTS OF DISCUSSION
%\textcolor{red}{
%\begin{itemize}
%	\item Summary of paper and main results
%	\item Analysis of results and comparison to state of the art but more of an emphasis on why the method produced those results or why the state of the art is limited
%	\item Expansion on certain results from experiments and what it actually means
%	\item Limitations and Future work
%	\item Conclusion
%\end{itemize}
%}
%\textcolor{blue}{The discussion is where you pull your results together into a coherent story, and put that story in context by referring back to your own results and to other peoples’ research. By the end of the discussion, you should have addressed the goals and objectives you outlined in your introduction. Look at other papers and think about what could potentiall go in here. Danny's NIMG paper is a good example. }  

We introduce a probabilistic deep learning (DL) framework for quantifying three types of uncertainties that arise in data-enhancement applications, and demonstrate its potential benefits in improving the safety of such systems towards practical deployment. The framework models \textit{intrinsic uncertainty} through heteroscedastic noise model and \textit{parameter uncertainty} through approximate Bayesian inference in the form of variational dropout, and finally integrates the two to quantify \textit{predictive uncertainty} over the system output.  Experiments focus on the super-resolution application of image quality transfer (IQT)\cite{alexander2017image} and study several desirable properties of such framework, which lack in the existing body of data enhancement methods based on deterministic DL models. 

%Firstly, results on a range of applications and datasets illustrate the benefits of uncertainty modelling on the accuracy and robustness of super-resolution algorithms.
Firstly, results on a range of applications and datasets show that modelling uncertainty improves overall prediction performance. Table~\ref{tab:compare_1} and \ref{tab:compare_2} show that modelling the combination of both \textit{intrinsic} and \textit{parameter} uncertainty achieves the state-of-the-art accuracy on super-resolution of DTIs and MAP-MRI coefficients in both of the HCP test dataset and the Lifespan dataset, improving on the present best methods based on random-forests (RF-IQT\cite{alexander2017image} and RF-BIQT\cite{tanno2016bayesian}) and interpolation---the standard method to estimate sub-voxel information used in clinical visualisation software. In particular, results on the Lifespan dataset, which differs from the training data in age range and acquisition protocol, indicates the better generalizability of our method. In addition, Fig.~\ref{fig:tract} shows that such combined model also benefits downstream tractography in comparison with the previous methods, illustrating the potential utility of the method for downstream connectivity analysis. Such improvement in the predictive performance arises from the regularisation effects imparted by the modelling of respective uncertainty components. Specifically, modelling intrinsic uncertainty through the heteroscedastic network improves robustness to outliers, while modelling parameter uncertainty via variational dropout defends against overfitting. For example,  Table~\ref{tab:compare_2} shows that the predictive performance of the 3D-ESPCN + Hetero. model is only marginally compromised even when the outliers are not removed from training data, while the baseline 3D-ESPCN results in much poorer performance. This can be ascribed to the ability of the variance network $\Sigma_{\theta_2}(\cdot)$ in the 3D-ESPCN + Hetero. architecture to attenuate the effects of outliers by assigning small weights (i.e. high uncertainty) in the weighted MSE loss function as shown in eq.~\eqref{eq:mc_intrinsic_uncertainty}. However, this loss attenuation mechanism can also encourage the network to overfit to low-uncertainty regions, potentially focusing less on ambiguous yet important parts of the data---we indeed observe in Table~\ref{tab:compare_2}  that the heteroscedastic network performs considerably worse than the baseline 3D-ESPCN on the exterior regions while the reverse is observed on the interior part. Such overfitting to low-uncertainty interior regions is alleviated by modelling parameter uncertainty with variational dropout \cite{kingma2015variational}, as evidenced by the dramatic error reduction in the exterior region on both HCP and Lifespan datasets.

 Secondly, experiments on the images of healthy and pathological brains have demonstrated the utility of \textit{predictive uncertainty} as a reliability metric of output images.  Fig.~\ref{fig:uncertainty_map} illustrates the strong correspondence between the maps of predictive uncertainty and the reconstruction quality (voxel-wise RMSE) in the downstream derived quantites such as FA and MD maps. In addition, Fig.~\ref{fig:components_pathology} shows that such uncertainty measure also highlights pathological structures not observed in the training data. We have also tested the utility of predictive uncertainty in discriminating voxels with sufficiently low RMSEs in the predicted high-resolution MD maps. As shown in Fig.~\ref{fig:roc}, the optimal threshold selected on the HCP training dataset is capable to detecting over $90\%$ of non-reliable predictions---voxels with RMSE above a certain threshold---not only on the unseen subjects in the same HCP cohort but also on subjects from the out-of-sample Lifespan dataset, that are statistically disparate from the training distribution (e.g. different age range and acquisition protocol). These results combined demonstrate the utility of predictive uncertainty map as a means to quantify output safety, and provides a subject-specific alternative to standard population-group reliability metrics (e.g. mean reconstruction accuracy in a held-out cohort of subjects). Such conventional group statistics can be misleading in practice; for instance, the information that a super-resolution algorithm is reliable $99\%$ of the time on a dataset of $1000$ subjects may not accurately represent the performance on a new unseen individual if the person is not well-represented in the cohort (e.g. pathology, different scanners, etc). In contrast, predictive uncertainty provides a metric of reliability, tailored to each individual at hand.

% Typically, the reliability of a super-resolution or any other data-enhancement algorithm is measured by population-group statistics such as reconstruction accuracy or performance in downstream tasks. However, this could be misleading in practice. For instance, the information that a super-resolution algorithm is reliable $99\%$ of the time on dataset $X$ consisting of $1000$ subjects is not useful if the new subject under consideration is not well-represented in dataset $X$ (e.g. pathology, different scanners, etc).  In contrast, these results show that predictive uncertainty provides a metric of reliability, tailored to each subject data at hand. 
  
 Thirdly, our preliminary experiments show that decomposition of the effects of intrinsic and parameter uncertainty in the predictive uncertainty provides a layer of explanations into the performance of the considered deep learning methods. Fig.~\ref{fig:healthy_abnormal} shows that the low reconstruction error in the centre of the benign cyst can be explained by the dominant intrinsic uncertainty, which indicates the inherent simplicity of super-resolution task in such homogeneous region, whilst the unfamiliarity of such structure in the healthy training dataset is reflected in the high parameter uncertainty. Assuming that the estimates of decomposed uncertainty components are sufficiently accurate, we could act on them to further improve the overall safety of the system. Imagine a scenario where reconstruction error is consistently high on certain image structures, if the parameter uncertainty is high but intrinsic uncertainty is low, this indicates that collecting more training data would be beneficial. On the other hand, if the parameter uncertainty is low and intrinsic uncertainty is high, this would mean that we need to regard such errors as inevitability, and abstain from predictions to ensure safety or account for them appropriately in subsequent analysis.
 
 The proposed methods for estimating intrinsic and parameter uncertainty, however, make several simplifying assumptions in the forms of likelihood model $p(\mathbf{y}|\theta, \mathbf{x})$ and posterior distributions over network parameters $p(\theta|\mathcal{D})$. Firstly, the likelihood model takes the form of a Gaussian distribution with a diagonal covariance matrix. This means that the likelihood model is not able to capture multi-modality of the predictive distribution i.e. the presence of multiple different solutions. While the full predictive distribution (eq.~\eqref{eq:full_distribution}) is not necessarily unimodal in theory due to the integration with the posterior distribution, we observe in practice that the drawn samples are not very diverse. Future work should explore the benefits of employing more complex forms of likelihood functions such as mixture models \cite{bishop1994mixture,kohl2018probabilistic}, diversity losses \cite{guzman2012multiple,bouchacourt2016disco,lee2018diverse} and more powerful density estimators \cite{huang2018multimodal,rezende2015variational,papamakarios2017masked,odena2017conditional,kohl2018probabilistic}. Also, the diagonality of covariance matrices means that the output pixels are assumed statistically independent given the input. Although the predicted images display high inter-pixel consistency, modelling the correlations between neighbouring pixels \cite{chandra2016fast} may further improve the reconstruction quality. Analogous to the likelihood function, variational dropout \cite{kingma2015variational}, which is used in this work, approximates the posteriors $p(\theta|\mathcal{D})$ by Gaussian distributions with diagonal covariance, imposing restrictive assumptions of unimodality and statistical independence between neural network weights. More recent advances in the Bayesian deep learning research \cite{louizos2016structured,oh2019radial,krueger2017bayesian,zhang2019cyclical,pawlowski2017implicit,louizos2017multiplicative} could be used to enhance the quality of parameter uncertainty estimation by allowing the model to capture multi-modality and statistical dependencies between parameters.  %\textcolor{red}{Would be interesting to talk about the mixing between intrinsic and parameter uncertainty. }

An important future challenge is the clinical validation of predictive uncertainty as a reliability metric of output images. To this end, we need to design a more clinically meaningful definition of success and failure of the data enhancement algorithm at hand. Despite the high accuracy in distinguishing between predictive failures and successes attained with our method (Fig.~\ref{fig:roc}),  our definition of reconstruction quality, namely voxel-wise RMSE, does not necessarily represent the real utility of the output image. One possible approach would be to have clinical experts to label the potential failures in the super-resolved images, be it for a targeted application (e.g. diagnosis of some neurological conditions) or for general usage in clinical practice. A more economical alternative, which does not require extra label acquisition, is to define the prediction success in downstream measurements of interest i.e. functions of the output images $g(\cdot)$, such as morphometric measurements of anatomical or pathological structures (e.g. volumes). The propagation method (eq.~\eqref{eq:variance_decomposition}) introduced in Sec.~\ref{sec:uncertainty_decom} can be utilised to quantify uncertainty components in the space of target measurement  $g(\cdot)$. Measuring the correlation between such propagated uncertainty estimates and the corresponding errors would be a useful indicator of how well the uncertainty measure reflects the accuracy of the chosen measurement $g(\cdot)$. Lastly, our initial results on the brain tumour dataset motivate a larger-scale quantitative validation of uncertainty estimates in the presence of pathology. Future work must examine the effect of including patients' dataset in the training data on the estimate of uncertainty components. 

There are many ways in which uncertainty information could be utilised by radiologists or other users of data enhancement algorithms. First, predictive uncertainty can be used to decide when to abstain from predictions in high-risk regions of images (e.g. anomalies, out-of-distribution examples or inherently ambiguous features). For example, the original input low-resolution image can be augmented by overlaying the high-resolution prediction only in locations with sufficiently low uncertainty, before presenting to clinicians. As demonstrated by Fig.~\ref{fig:roc} in the context of super-resolution, such uncertainty-based quality control of predictions is potentially an effective means to maintain high accuracy of output images and also to safeguard against hallucination or removal of structures \cite{cohen2018distribution}. Second, the uncertainty information could be used for active learning \cite{settles2009active} to decide which images should be labelled and included in the training set to maximally improve the model performance. Prior work \cite{gal2017deep,gorriz2017cost} define the acquisition function so as to select examples with high parameter uncertainty, and achieve promising results in classification and segmentation tasks. In particular, these methods are able to construct a compact and effective training dataset, and consequently improve the prediction accuracy while reducing the training time. The same idea could be naturally extended to data enhancement problems, that are typically formulated as multivariate regression tasks. For example, in the case of IQT, we could simulate a library of low-resolution and high-resolution image pairs from a large public dataset (e.g. HCP), and incrementally expand the training data by adding more examples from such a library. We should note, however, that in many data enhancement applications, obtaining a new ``label'' may require an extra acquisition possibly with a different scanner or modality, which may be logistically challenging. Third, another important application is transfer learning \cite{pan2010survey} where uncertainty information could be used to leverage knowledge from different but related domains or tasks. In many data enhancement applications, the test distribution can considerably deviate from the training distribution. For example, the algorithm might be trained on a synthetic dataset or images acquired from a scanner that is very different from the one used in the hospital where one plans to deploy the model. Therefore, a mechanism to adapt performance within a specific environment (e.g., based on the local patient population) \cite{kamnitsas2017unsupervised}, possibly in an online fashion \cite{karani2018lifelong,baweja2018towards}, is in demand. Recent work have shown that the Bayesian formalism provides a natural framework to use uncertainty in order to account for the difference and commonality between distributions to guide information transfer in continual learning \cite{kirkpatrick2017overcoming,nguyen2017variational} or few-shot learning \cite{finn2018probabilistic,yoon2018bayesian} settings. Exploring the benefits of these ideas in the context of medical image enhancement remains future work.

%Mention that the lack of paired data is hard to come by (with a few exceptions e.g. travelling head dataset -- can ask Stef to help). Also, the domain adaptation/transfer learning are important domains, and the use uncertainty information. Kosta's continual learning \cite{baweja2018towards} and domain adaptation paper \cite{kamnitsas2017unsupervised}. 
%Variational continual learning , elastic weight consolidation \cite{kirkpatrick2017overcoming}, lifelong learning \cite{karani2018lifelong}. 
%Approximate using the approximate posterior as the prior for the next domain 

The proposed framework for uncertainty quantification is formulated for multivariate regression in the general form, and thus is naturally applicable to many other image enhancement challenges such as: rapid image acquisition techniques e.g., compressed sensing \cite{sun2016deep}, MR fingerprinting \cite{ma2013magnetic,cohen2018mr} or sparse reconstruction \cite{schlemper2018deep,hammernik2018learning}; denoising \cite{benou2017ensemble} and dealiasing \cite{yang2018dagan,han2018deep}; image synthesis tasks e.g., estimating T2-weighted images from T1 \cite{rousseau2008brain,ye2013modality,jog2015mr}, estimating CT images from MRI \cite{burgos2015robust,bragman2018uncertainty,nie2018medical}, and generating a high-field scan from a low-field scan \cite{bahrami2016convolutional}; data harmonisation
\cite{mirzaalian2016inter,karayumak2018harmonizing,tax2019cross} which aims to learn mappings among imaging protocols to reduce confounds in multicentre studies. Our results on image quality transfer \cite{alexander2017image} illustrate the potential of the uncertainty modelling techniques to improve the safety of these applications by not only improving the predictive accuracy, but also providing a mechanism to quantify risks and safeguard against potential malfunction. 

\clearpage
\section*{Data and code availability statement}
The Human Connectome Project dataset (release Q3) \cite{van2013wu} and the Lifespan dataset \cite{harms2018extending} are publicly available. The Prisma data is available upon request. The glioma and multiple-sclerosis datasets are part of on-going studies at the Humanitas Research Hospital, Italy and Institute of Neurology at UCL, UK repectively, and we are bound by the policies of the data providers. The code will be released at \url{https://github.com/rtanno21609/SaferNeuroimageEnhancement} upon publication.

\section*{Acknowledgment}
We would like to thank Felix Bragmann at Babylon Health, Zach Eaton-Rosen at UCL/KCL and Stefano Blumberg at UCL for their valuable feedback. We would also like to thank Samuel Hurley whom helped with the Prisma acquisitions in FMRIB at University of Oxford. The tumour data were acquired as part of a clinical research project lead by Alberto Bizzi, MD at the Humanitas Research Hospital in Milan, Italy. We are also grateful to Mark S. Graham at Visulytix and Gary Zhang at UCL for connecting us with Alberto Bizzi. The multiple sclerosis (MS) data were acquired as part of a study at UCL Institute of Neurology, funded by the MS Society UK and the UCL Hospitals Biomedical Research Centre (PIs: David Miller and Declan Chard). The HCP data were provided by the WU-Minn Consortium (PIs: David Van Essen and Kamil Ugurbil; 1U54MH091657) funded by NIH and Washington University.

 EU Horizon 2020 grant CDS-QuaMRI 634541-2 and EPSRC grants R014019 R006032 N018702 and M020533 support DCA’s work on this topic. FG has received funding under the European Union’s Horizon 2020 research and innovation programme under grant agreement No. 634541 and from the EPSRC (R006032/1, M020533/1). RT was supported by Microsoft scholarship.



%\bibliographystyle{IEEEtran}
\bibliographystyle{IEEEtran}
\bibliography{reference}

\end{document}